\documentclass[twocolumn]{aastex63}
\hypersetup{linkcolor=blue,citecolor=blue,filecolor=cyan,urlcolor=blue}
\usepackage{ulem}
\usepackage{lineno}

\submitjournal{ApJ}

\shorttitle{HD 106906}
\shortauthors{Crotts et al.}
\graphicspath{{./}{figures/}}

\begin{document}

\title{A Deep Polarimetric Study of the Asymmetrical Debris Disk HD 106906}

\correspondingauthor{Katie Crotts}
\email{ktcrotts@uvic.ca}

\author[0000-0003-4909-256X]{Katie A. Crotts}
\affiliation{University of Victoria, 3800 Finnerty Rd, Victoria, BC, V8P 5C2, Canada}

\author[0000-0003-3017-9577]{Brenda C. Matthews}
\affiliation{University of Victoria, 3800 Finnerty Rd, Victoria, BC, V8P 5C2, Canada}
\affiliation{National Research Council of Canada Herzberg, 5071 West Saanich Road, Victoria, BC V9E 2E7, Canada}

\author[0000-0002-0792-3719]{Thomas M. Esposito}
\affiliation{Astronomy Department, University of California, Berkeley, CA 94720, USA}
\affiliation{SETI Institute, Carl Sagan Center, 189 Bernardo Ave., Mountain View CA 94043, USA}

\author[0000-0002-5092-6464]{Gaspard Duch\^{e}ne}
\affiliation{Astronomy Department, University of California, Berkeley, CA 94720, USA}
\affiliation{Universit\'{e} Grenoble Alpes/CNRS, Institut de Plan\'{e}tologie et d'Astrophysique de Grenoble, 38000 Grenoble, France}

\author[0000-0002-6221-5360]{Paul Kalas}
\affiliation{Astronomy Department, University of California, Berkeley, CA 94720, USA}
\affiliation{SETI Institute, Carl Sagan Center, 189 Bernardo Ave., Mountain View CA 94043, USA}
\affiliation{Institute of Astrophysics, FORTH, GR-71110 Heraklion, Greece}

\author[0000-0002-8382-0447]{Christine H. Chen}
\affiliation{Space Telescope Science Institute (STScI), 3700 San Martin Drive, Baltimore, MD 21218, USA}

\author[0000-0001-6364-2834]{Pauline Arriaga}
\affiliation{Department of Physics and Astronomy, 430 Portola Plaza, University of California, Los Angeles, CA 90095, USA}

\author[0000-0001-6205-9233]{ Maxwell A. Millar-Blanchaer}
\altaffiliation{NASA Hubble Fellow}
\affiliation{NASA Jet Propulsion Laboratory, California Institute of Technology, Pasadena, CA 91109, USA}

\author[0000-0002-1783-8817]{John H. Debes}
\affiliation{Space Telescope Science Institute (STScI), 3700 San Martin Drive, Baltimore, MD 21218, USA}

\author[0000-0002-1834-3496]{Zachary H. Draper}
\affiliation{University of Victoria, 3800 Finnerty Rd, Victoria, BC, V8P 5C2, Canada}
\affiliation{National Research Council of Canada Herzberg, 5071 West Saanich Road, Victoria, BC V9E 2E7, Canada}

\author[0000-0002-0176-8973]{Michael P. Fitzgerald}
\affiliation{Department of Physics and Astronomy, 430 Portola Plaza, University of California, Los Angeles, CA 90095, USA}

\author[0000-0001-9994-2142]{Justin Hom}
\affiliation{School of Earth and Space Exploration, Arizona State University, PO Box 871404, Tempe, AZ 85287, USA}

\author{Meredith A. MacGregor}
\altaffiliation{NSF Astronomy and Astrophysics Postdoctoral Fellow}
\affiliation{Department of Terrestrial Magnetism, Carnegie Institution for Science, 5421 Broad Branch Road, Washington, DC 20015, USA}
\affiliation{Department of Astrophysical and Planetary Sciences, University of Colorado, 2000 Colorado Avenue, Boulder, CO 80309, USA}

\author[0000-0002-9133-3091]{Johan Mazoyer}
\altaffiliation{NFHP Sagan Fellow}
\affiliation{LESIA, Observatoire de Paris, Universit\'{e} PSL, CNRS, Sorbonne Universit\'{e}, Universit\'{e} de Paris, 5 place Jules Janssen, 92195 Meudon, France}

\author{Jennifer Patience}
\affiliation{School of Earth and Space Exploration, Arizona State University, PO Box 871404, Tempe, AZ 85287, USA}

\author[0000-0002-7670-670X]{Malena Rice}
\affiliation{Department of Astronomy, Yale University, New Haven, CT 06511, USA}

\author[0000-0001-6654-7859]{Alycia J. Weinberger}
\affiliation{Department of Terrestrial Magnetism, Carnegie Institution for Science, 5421 Broad Branch Road, Washington, DC 20015, USA}

\author[0000-0003-1526-7587]{David J. Wilner}
\affiliation{Center for Astrophysics, Harvard \& Smithsonian, 60 Garden St., Cambridge, MA 02138, USA}

\author[0000-0002-9977-8255]{Schuyler Wolff}
\affiliation{Department of Astronomy, The University of Arizona, 933 N Cherry Ave, Tucson, AZ, 85750, USA}



\begin{abstract}

HD 106906 is a young, binary stellar system, located in the Lower Centaurus Crux (LCC) group. This system is unique among discovered systems in that it contains an asymmetrical debris disk, as well as an 11 M$_{\text{Jup}}$ planet companion, at a separation of $\sim$735 AU. Only a handful of other systems are known to contain both a disk and directly imaged planet, where HD 106906 is the only one in which the planet has apparently been scattered. The debris disk is nearly edge on, and extends roughly to $>$500 AU, where previous studies with HST have shown the outer regions to have high asymmetry. To better understand the structure and composition of the disk, we have performed a deep polarimetric study of HD 106906’s asymmetrical debris disk using newly obtained $H$-, $J$-, and $K1$-band polarimetric data from the Gemini Planet Imager (GPI). An empirical analysis of our data supports a disk that is asymmetrical in surface brightness and structure, where fitting an inclined ring model to the disk spine suggests that the disk may be highly eccentric ($e\gtrsim0.16$). A comparison of the disk flux with the stellar flux in each band suggests a blue color that also does not significantly vary across the disk. We discuss these results in terms of possible sources of asymmetry, where we find that dynamical interaction with the planet companion, HD 106906b, is a likely candidate.

\end{abstract}



\section{Introduction} \label{sec:level1}
Debris disks can be characterized as gas-poor, dusty disks around main sequence stars, and which are important objects to study in relation to better understanding the evolution of exoplanetary systems. Debris disks are formed through collisional cascade, in which colliding planetismals, such as comets and asteroids, break into milli-meter to sub-micron sized dust grains (\citealt{Hughes:2018aa}; \citealt{Matthews:2014aa}; \citealt{Wyatt:2008aa}). For this reason, debris disks are indicators of the successful formation of at least $>$1 meter-sized objects during the protoplanetary disk phase. However, processes from the central star(s), such as radiation pressure and stellar winds, cause dust grains under a certain size to be blown out from the system (\citealt{Augereau:aa}; \citealt{Strubbe:2006aa}). Therefore, in order for a collisional cascade to occur and continually produce smaller dust grains to replenish the disk, planetesimals within the system need to have a sufficient relative velocity to cause fragmentation upon impact, meaning that some sort of stirring mechanism is required to perturb their orbits. 

There are several stirring mechanisms that could occur, such as late planetismal formation at large distances from the star (\citealt{Kennedy:2010aa}; \citealt{Kenyon:2008aa}), interaction with the interstellar medium (ISM, \citealt{Debes:2009aa}), close encounters with nearby stars (\citealt{Kennedy:2014aa}), and planetary scattering (\citealt{Kalas:2015aa}). The source of stirring may be determined by direct observation, as each of these mechanisms can affect the morphology and properties of the disk in slightly different ways, such as creating warps, eccentric disks, gaps, and rings. In any of these cases, the chances of creating an asymmetrical debris disk in terms of brightness and structure are high. For this reason, asymmetrical debris disks are especially interesting and make great candidates for study as they allow us to gain more insight into the architecture of exoplanetary systems.

The Gemini Planet Imager (GPI, \citealt{Macintosh:2014aa}) has not only searched for exoplanets, but has imaged many debris disks as well. From November 2014 to October 2018, GPI conducted its Exoplanet Survey (GPIES, PI B. Macintosh, \citealt{Macintosh:2018aa}), which was an 860-hour campaign given to search roughly 600 stars for Jupiter-type planets, as well as debris disks observable through scattered light. Through this campaign, 26 debris disk targets were resolved via total intensity and/or polarized scattered light (\citealt{Esposito:2020aa}).

One of these disks is HD 106906 (HIP 59960), which was first discovered by Spitzer through an infrared survey of 25 stars in the Lower Centaurus Crux group (\citealt{chen05}) and is located $\sim$103.3 pc away (\citealt{Collaboration:2018aa}). HD 106906 is a unique and interesting system for several reasons. First, it contains an eccentric, tight binary (e=0.669$\pm$0.002), with two F5V type stars (\citealt{Rosa:2019aa}). Secondly, it has a possibly scattered 11$\pm$2 M$_{\text{Jup}}$ planet companion outside of the disk, at a separation of 735$\pm$5 AU (\citealt{Bailey:2013aa}; \citealt{Daemgen:2017aa}). And finally, the HD 106906 system hosts a debris disk that features a brightness asymmetry, with a south-east (SE) extension that appears brighter than the north-west (NW) extension (\citealt{Kalas:2015aa}; \citealt{Lagrange:2016aa}). 

The Hubble Space Telescope (HST), which has a much larger field of view than GPI, has also taken images of this system at optical wavelengths. The results from HST have shown that the NW side of the disk halo seems to extend to $>$500 AU, creating a ``needle" like structure (\citealt{Kalas:2015aa}). On the other hand, this same feature is not observed for the SE-extension. The HST data also show that the planet has a position angle (PA) $\sim$21 degrees from the NW-extension. Because of the wide separation of HD 106906b, its orbit is challenging to constrain; however, recent analysis has presented the first detection of the planet's orbit. If the planet is bound, this analysis shows a highly eccentric, non-coplanar orbit with a periastron distance that comes reasonably close to the disk (\citealt{Nguyen2020}). 

The fact that HD 106906 hosts an asymmetric debris disk, as well as an planet companion external to the disk, raises questions about whether or not HD 106906b is responsible for this asymmetry. Previous work has shown that if HD 106906b indeed formed within the disk, it could have migrated inwards, and consequently been ejected through dynamical interactions with the stellar binary (\citealt{Rodet:2017aa}). While an in situ formation scenario for the planet is possible, this is less favored compared to the scattering scenario due to the lack of circumstellar gas at that distance, as well as the planet formation timescale being greater than the lifespan of gas in a fiducial protoplanetary disk. In both formation scenarios, however, secular effects from the orbiting planet could create major asymmetries in the disk (\citealt{Rodet:2017aa}; \citealt{Nesvold:2017aa}). If the scattering scenario were true, the planet may have also had dynamical interactions with the disk on its scattered path, leading to a perturbed disk. 

The goal of this paper is to better understand the source of the disk's asymmetry. To do this, we analyzed the latest GPI data, revealing evidence of a geometrically asymmetric and eccentric disk. We further studied the disk through 3D radiative-transfer modelling, gaining a deeper look at the disk's density structure and dust grain properties. We chose to concentrate on the polarimetric data of HD 106906 debris disk, as polarimetry allows us to better probe the dust grains and gives a better representation of the total disk structure. We discuss our results in terms of dynamical perturbation from HD 106906b and an ISM interaction.

\section{Observations and Data Reduction} \label{sec:level2}

\subsection{GPI Observations}

We obtained deep polarimetric data of HD 106906 in the $H$-band (peak $\lambda=1.647$ $\mu$m) from GPIES. This data has a greater total integration time (2564.79s compared to 709.94s) and field rotation (20.3$^{\circ}$ compared to 7.1$^{\circ}$) than the polarimetric data presented in Kalas et al. (\citeyear{Kalas:2015aa}). We also obtained polarimetric $J$- (peak $\lambda=1.235$ $\mu$m) and $K1$-band (peak $\lambda=2.048$ $\mu$m) data through one of Gemini Observatory's Large and Long Programs (LLP, PI Christine Chen). The $H$-band data was taken June 30$^{\text{th}}$, 2015, using GPI's polarimetric (H-Pol) mode. The $J$- and $K1$-band data were later taken over a four day period from March 24$^{\text{th}}$ through the 28$^{\text{th}}$ of 2016, again in polarimetric (J-Pol and K1-Pol) mode. Due to a malfunction in GPI's apodizer wheel at that time, all three observations used the coronagraph apodizing mask that was optimized for $H$-band (``APOD H G6205''). Further details about the polarimetric observations can be found in Table~\ref{tab:table1}. 

\begin{table*}[ht!]
\caption{\label{tab:table1}%
Details of observations with GPI for each band. t$_{\text{exp}}$ = the exposure time for each frame, N = the number of frames, $\Delta$PA = the total Parallactic Angle rotation.
}
\begin{ruledtabular}
\begin{tabular}{ccccccc}
\multicolumn{1}{c}{\textrm{Mode}}&\textrm{Band}&\textrm{Date}&\textrm{t$_{\text{exp}}$ (s)}&\textrm{N}&\textrm{$\Delta$PA (deg)}&\textrm{MASS Seeing}\\
\colrule
Pol & J & 2016 Mar 26 & 59.65 & 54 & 35.2 & 0.77$''$ \\
Pol & H & 2015 Jun 30 & 59.65 & 43 & 20.3 & 0.27$''$  \\
Pol & K1 & 2016 Mar 28 & 88.74 & 40 & 36.5 & 1.33$''$ \\
\end{tabular}
\end{ruledtabular}
\end{table*}

\subsection{Data Reduction}
Each of these sets of data were reduced through the GPI Data Reduction Pipeline (\citealt{Perrin:2014aa}). In GPI's polarization mode, a pair of spots is created by each lenslet, representing both orthogonal polarization states Q and U. The first step in the reduction process is to create 3-dimensional cubes from the raw data, where the three dimensions contain an image for each orthogonal polarization state. The raw data are dark subtracted, bad pixel corrected and cleaned of correlated detector noise, then combined into polarization data cubes. The images are then corrected for geometric distortion. The position of the star is found by measuring a set of fiducial satellite spots (\citealt{Wang:2014aa}). 

Next, the data cubes are combined to create the final composite data cube. In this final step, the instrumental polarization is measured from the apparent stellar polarization in each datacube by measuring the average fractional polarization inside the occulting mask, which is then subtracted from each cube (\citealt{Millar-Blanchaer:2015aa}). Additionally, each cube is smoothed with a Gaussian kernel (FWHM of 1 pixel) and the satellite-spot-to-star flux is measured in order to determine a photometric calibration factor. After this measurement, the 3-D cubes are combined to create a single radial Stokes image, consisting of Stokes [I,Q$_{\phi}$,U$_{\phi}$,V]. Finally, using the stellar magnitude and photometric calibration factor for each filter+apodizer combination, the data were calibrated from ADU coadd$^{-1}$ to Jy arcsec$^{-2}$ \citep{Hung:2016}. 

\begin{figure*}
\centering
\includegraphics[scale=.61]{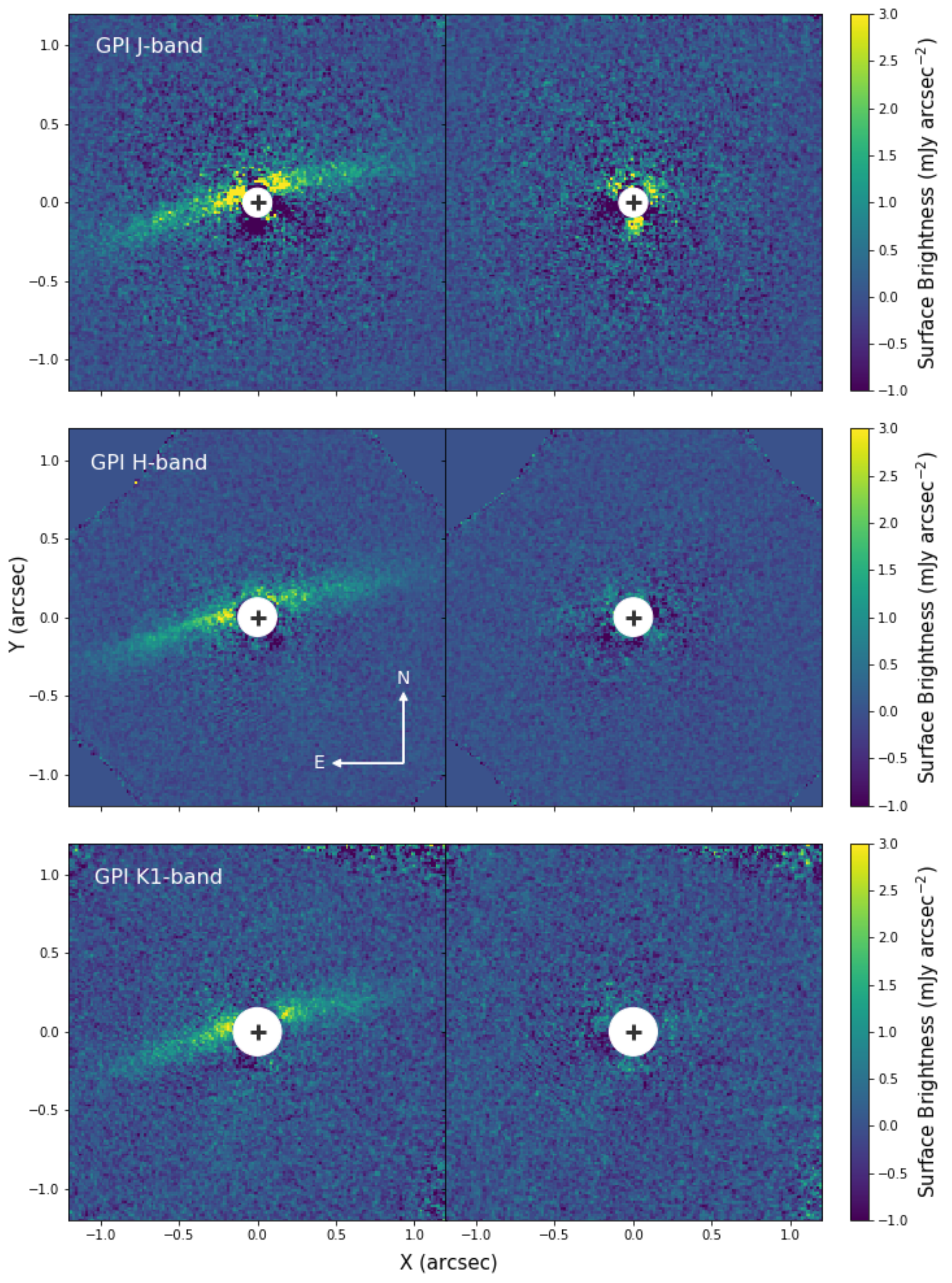}
\caption{\label{fig:fig1}The HD 106906 debris disk in $J$-, $H$- and $K1$-band polarized intensity as observed with GPI. \textbf{Left column} depicts the disk in Q$_{\phi}$. \textbf{Right column} depicts the disk in U$_{\phi}$. White circles denote regions obscured by GPI's FPM and black crosses mark the star location.}
\end{figure*}

The final images in each filter can be seen in Figure~\ref{fig:fig1}, where the left column represents the data in Q$_{\phi}$, and the right column represents the data in U$_{\phi}$. The white circles represent the size of the focal plane mask (FPM), with a radius of 0.09$''$, 0.12$''$ and 0.15$''$ respectively.

\graphicspath{{./}{Figures/}}

\section{Observational Results} \label{sec:level3}

\begin{figure*}
\centering
\includegraphics[scale=.8]{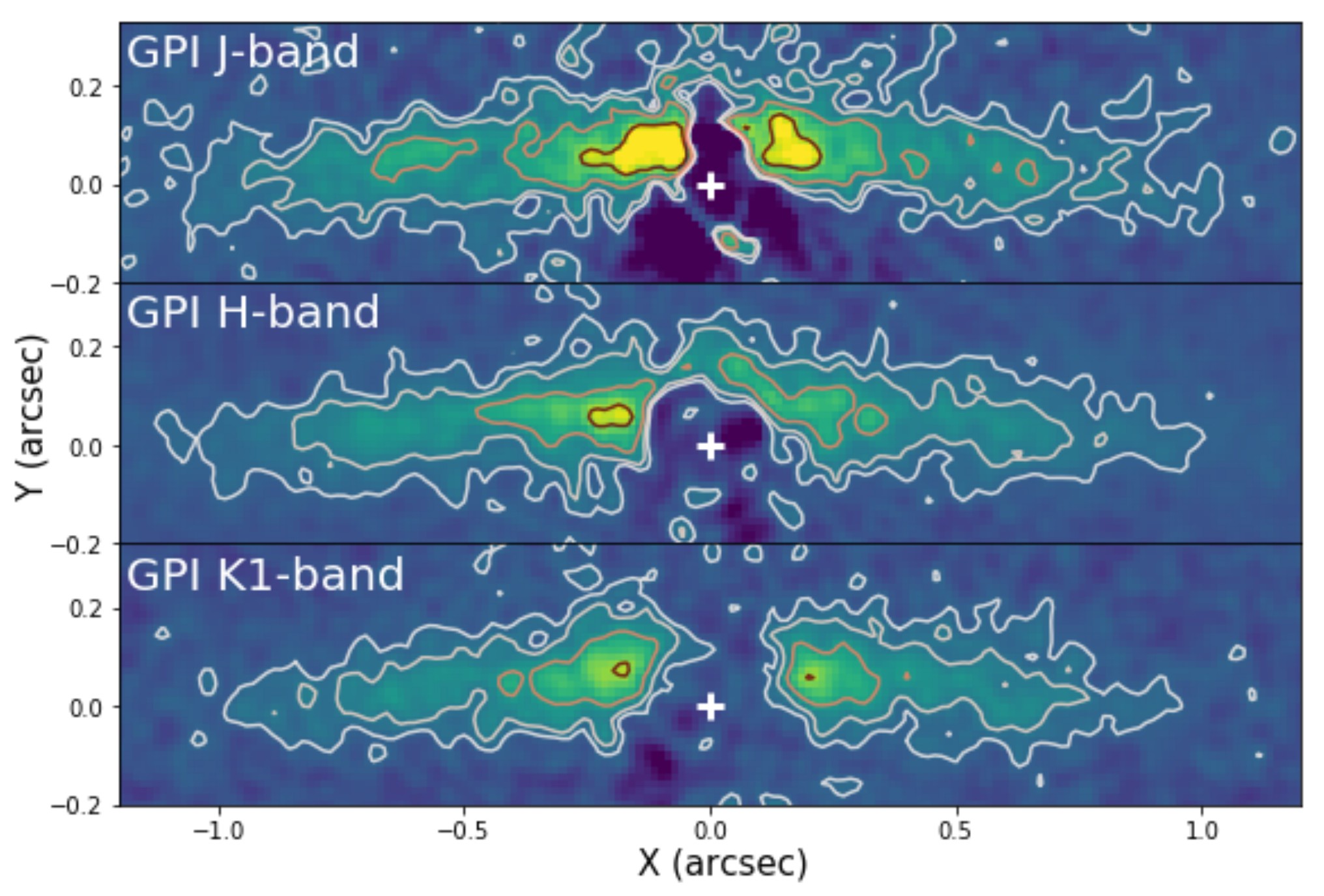}
\caption{\label{fig:fig14}$J$-, $H$- and $K1$-band polarimetric intensity, overlaid with surface brightness contours. Scaling is the same as in Figure~\ref{fig:fig1}. Contours represent a surface brightness of 0.3, 0.6, 1.2, and 2.4 mJy/arcsec$^{2}$, where the outer most contour corresponds with 2$\sigma$.}
\end{figure*}

Consistent with previous observations, we find the disk to be very forward scattering, where only the front side of the disk is visible. As seen in Figure~\ref{fig:fig1}, a brightness asymmetry can be visually observed in all three bands, where the SE-extension is brighter than the NW-extension. This asymmetry is most prominent in the $H$-band, although the asymmetry seems to be present in the $J$- and $K1$-band as well. Quantifying the surface brightness in these two bands will help confirm this. Figure~\ref{fig:fig14} shows a closer look at the disk in each band, smoothed with a Gaussian ($\sigma=1$ pixel), and overlaid with contours to accentuate the surface brightness features. 

\begin{figure*}
\centering
\includegraphics[width=0.95\textwidth]{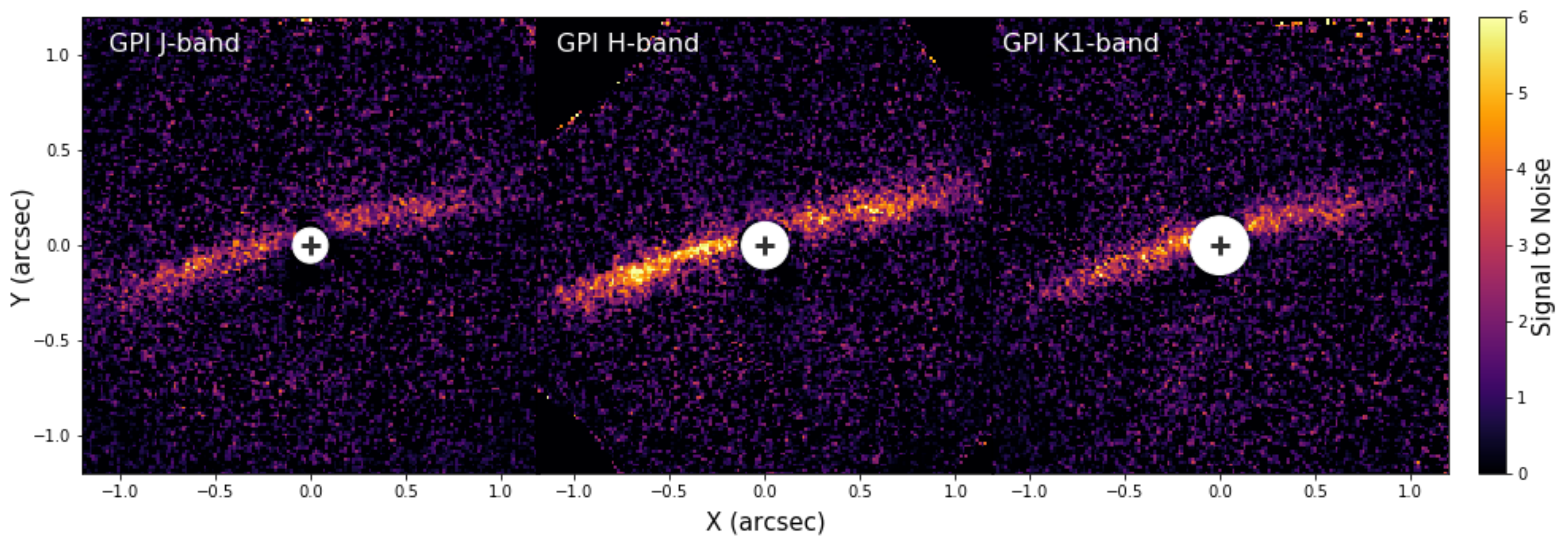}
\caption{\label{fig:fig2}Signal to Noise maps for HD 106906 in $J$-, $H$-, and $K1$-bands. Note that $H$-band has the highest signal to noise. White circles denote regions obscured by GPI's FPM and black crosses mark the star location.}
\end{figure*}

Additionally, we created noise maps at each wavelength from the U$_{\phi}$ image as the standard deviation at each radius in 1-pixel wide stellocentric annuli. As expected for an optically thin debris disk causing single scatterings, we assume the corresponding U$_{\phi}$ images contain noise but little or no disk signal, although we acknowledge a small amount of disk signal could be present due to smearing introduced by the finite size of the PSF (\citealt{Heikamp_19}). By dividing these noise maps from our Q$_{\phi}$ images, we create a Signal to Noise (S/N) image for each wavelength, which can be seen in Figure~\ref{fig:fig2}. Comparing the S/N between each band, we find that the $H$-band has the highest S/N, while $J$-band has the lowest S/N, likely due to lower Strehl and poorer seeing. Using a S/N of 2$\sigma$ (corresponding with the outermost contour in Figure~\ref{fig:fig14}), we find that the disk extends roughly between 1.0$''$ and 1.1$''$ (103-114 AU) for the three bands, where we do not observe a significant radial extent asymmetry. This is consistent with results found for the polarized data analyzed in Kalas et al. (\citeyear{Kalas:2015aa}), where no radial extent asymmetry was found in contrast to their total intensity data.  

\subsection{Surface Brightness} \label{sec:flux}
\begin{figure}
\includegraphics[scale=.319]{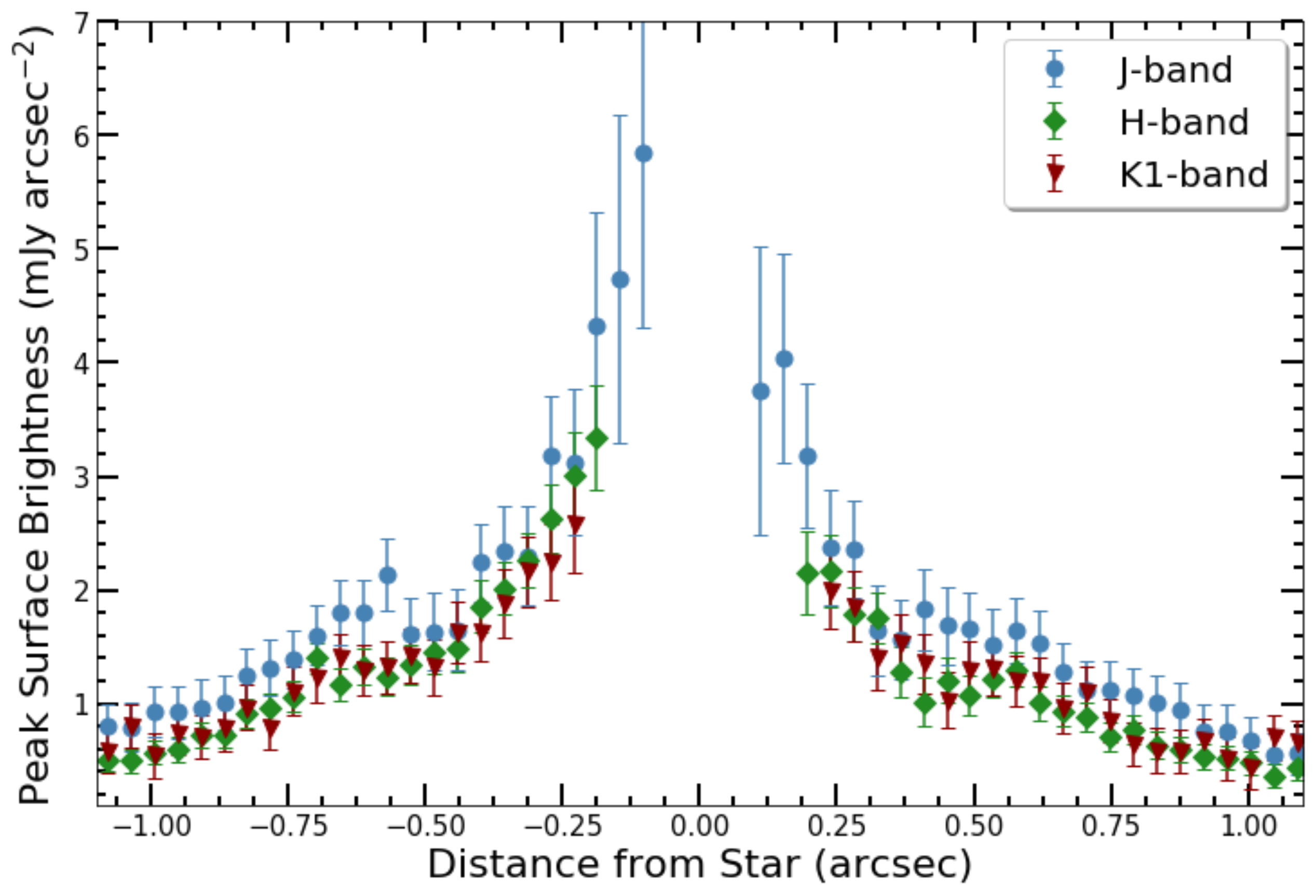}
\caption{\label{fig:fig4}Peak surface brightness in each band as a function of radial separation from the star in arcseconds.}
\end{figure}

For this section, we calculated the peak surface brightness profile in each wavelength in order to quantify the surface brightness asymmetry. To do this, we find the peak brightness along the spine of the disk. After rotating the disk to be horizontal, we find the maximum (``peak'') surface brightness along vertical slices. The peak surface brightness is then binned by 3-pixels along the horizontal axis and averaged. These results can be found in Figure~\ref{fig:fig4}, showing the peak surface brightness in mJy arcsec$^{-2}$ as a function of projected separation from the star in arcseconds. The error bars are calculated using the noise map in each filter. 

In all three bands, we observe the same general shape of the surface brightness profiles. Within $\sim$0.50$''$, the surface brightness peaks closest to the star and decreases steeply towards larger stellar separations. Approximately between 0.50$''$ and 0.70$''$, the surface brightness briefly plateaus before again decreasing more gradually out to $\sim$1.1$''$.

Looking at the relative peak surface brightness between each band, we confirm that there is a brightness asymmetry in all three wavelengths across the disk. Table~\ref{tab:asymm} shows the percentage by which the SE-extension is brighter than the NW-extension, segmented by radial separation from the star. While the asymmetry differs slightly between each band, all three exhibit a SE-extension that is roughly 10-30\% brighter than the NW-extension. We find that the $H$-band exhibits the greatest asymmetry with 30.9$\pm$4.4\% brighter flux in the SE-extension between the inner working angle (IWA) and 0.40$''$, about 10\% greater than that measured in Kalas et al. (\citeyear{Kalas:2015aa}), and shows another peak of 27.0$\pm$3.9\% between 0.60$''$ and 0.80$''$. The asymmetries in the $J$- and $K1$-bands are more modest, with a smaller asymmetry along the disk compared to the $H$-band. While the $J$- and $K1$-band have larger uncertainties, because these three bands are very close in wavelength, we do not expect that the true surface brightness asymmetry varies too significantly between them. The asymmetry seen in the $H$-band is more likely a better representation of the true brightness asymmetry observed at these near-IR wavelengths, due to its much higher S/N. Regardless, this analysis shows that a modest surface brightness asymmetry exists along the disk in all three bands, and that the previously resolved asymmetry does not result from an artifact.

\begin{figure}
\includegraphics[scale=.319]{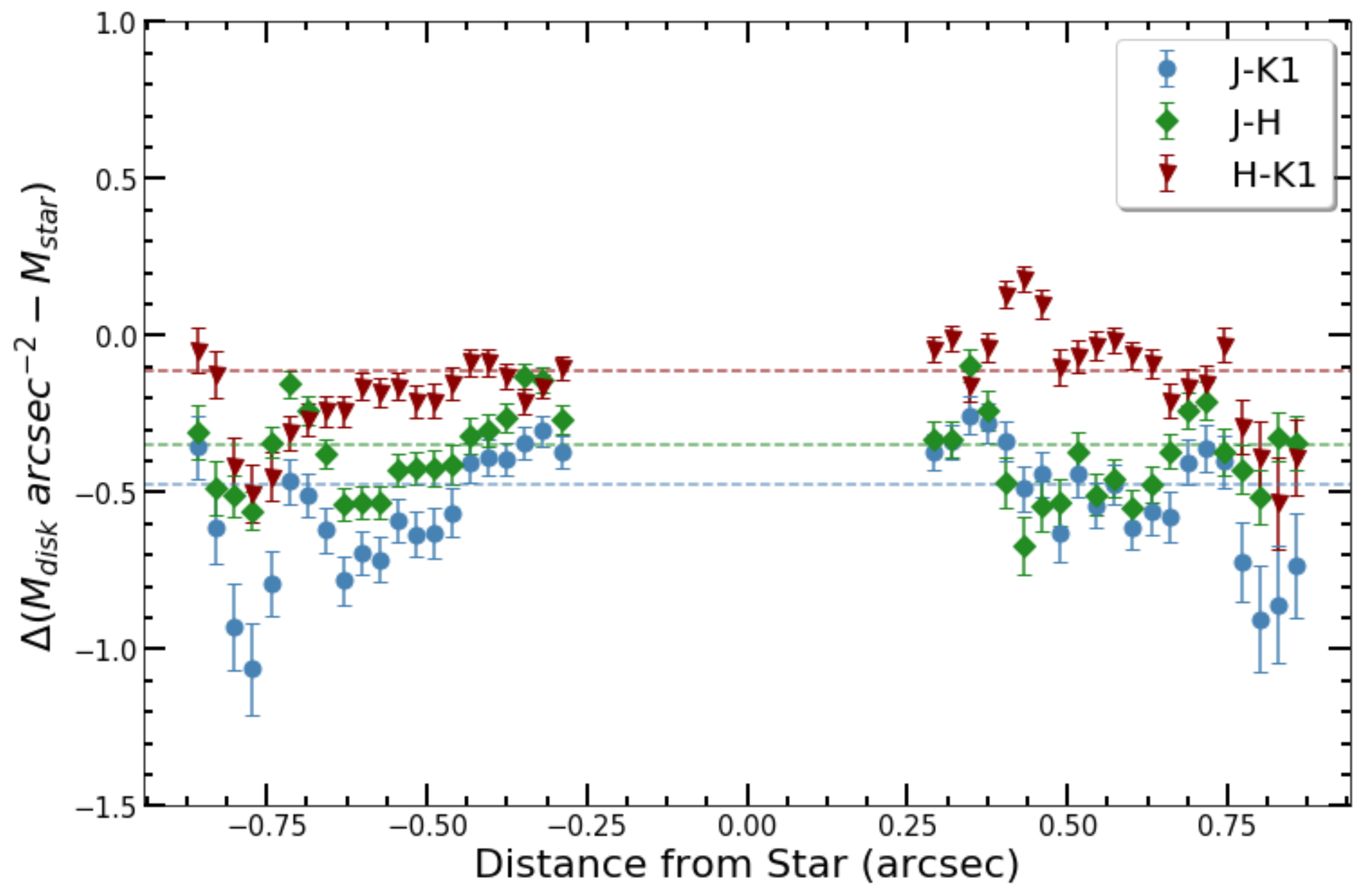}
\caption{\label{fig:disk_color}Disk color between each band as a function of radial separation from the star in arcseconds. Horizontal dashed lines represent the weighted means for $J$-$K1$ (blue), $J$-$H$ (green), and $H$-$K1$ (red).}
\end{figure}

Comparing the flux of the disk to the flux of the star in different bands tells us about the disk scattered-light color. Figure~\ref{fig:disk_color} shows the disk color as a function of the difference in the disk surface brightness minus the stellar magnitude between each band. We use the 2MASS $J$-, $H$-, and $K$-band magnitudes for HD 106906AB (6.946$\pm$0.03 mag, 6.759$\pm$0.038 mag and 6.683$\pm$0.026 mag respectively). To measure the color, the disk flux in each band is integrated between 0.30$''$-0.85$''$ and over $\sim$0.15$''$ vertically on both sides of the disk. The total flux is then converted to magnitudes arcsec$^{-2}$ and compared between two bands, where the stellar magnitudes in each band are then subtracted (i.e. $\Delta$(M$_{\text{disk}}$ arcsec$^{-2}$ - M$_{\text{star}}$) for $H$-$K1$ = $(H-K1)_{disk} - (H-K1)_{star}$). Here, a value above 0 indicates a red disk color, a value below 0 indicates a blue disk color, and a value of 0 indicates a neutral disk color. We find that the disk exhibits predominantly a blue color, where calculating $\Delta$(M$_{\text{disk}}$ arcsec$^{-2}$ - M$_{\text{star}}$) between each band gives $J$-$H$=-0.38$\pm$0.06 mag, $J$-$K1$=-0.52$\pm$0.07 mag, and $H$-$K1$=-0.14$\pm$0.05 mag. The difference between $J$-$K1$ and $H$-$K1$ also suggests that the disk color becomes increasingly more neutral towards longer wavelengths, consistent with disk color trends seen with other debris disks such as HD 15115 and HR 4796A (\citealt{Rodigas:2012aa}; \citealt{Rodigas2015}).

The disk color gives some information about the dust grains in the disk, as color is associated with certain dust grain properties such as the minimum dust grain size and porosity. For example, at a wavelength of 1.6 $\mu$m ($\sim$\textit{H}-band) versus 2.2 $\mu$m ($\sim$\textit{K}-band) and a porosity of 0, numerical simulations have shown that a strong blue disk color could be produced by a sub-micron minimum dust grain size, whereas a neutral disk color could be produced by a larger minimum dust grain size of several microns. A porosity greater than 0 can also increase the estimated minimum dust grain size as well (\citealt{Boccaletti:2003aa}; \citealt{Kirchschlager:2013aa}). The fact that our $H$-$K1$ color measurement is only weakly blue and close to neutral suggests the disk may have a larger minimum dust grain size. This being said, the minimum dust grain size is difficult to untangle and quantify from the disk color alone as dust composition and the scattering phase function (SPF) also have an effect. What is more important, is that we do not see a difference in color between the two extensions, implying that the dust grain properties do not significantly vary on either side of the disk.

\begin{table*}
\caption{\label{tab:asymm}%
Brightness asymmetry between the NW- and SE-extensions for each band as a function of stellar separation. Units are in percentage of how much brighter the SE-extension is compared to the NW-extension.
}
\begin{ruledtabular}
\begin{tabular}{ccccc}
\multicolumn{1}{c}{\textrm{Band}}&\textrm{IWA-0.40$''$}&\textrm{0.40$''$-0.60$''$}&\textrm{0.60$''$-0.80$''$}&\textrm{0.80$''$-1.0$''$}\\
\colrule
J & 26.9$\pm$9.7\% & 9.7$\pm$12.6\% & 22.3$\pm$10.7\% & 18.3$\pm$15.0\% \\
H & 30.9$\pm$4.4\% & 18.9$\pm$4.1\% & 27.0$\pm$3.9\% & 21.0$\pm$3.9\% \\
K1 & 23.4$\pm$11.2\% & 14.5$\pm$11.2\% & 18.0$\pm$12.3\% & 23.9$\pm$17.2\% \\
\end{tabular}
\end{ruledtabular}
\end{table*}

\subsection{Vertical Height} \label{sec:fwhm}
\begin{figure}
\includegraphics[scale=.32]{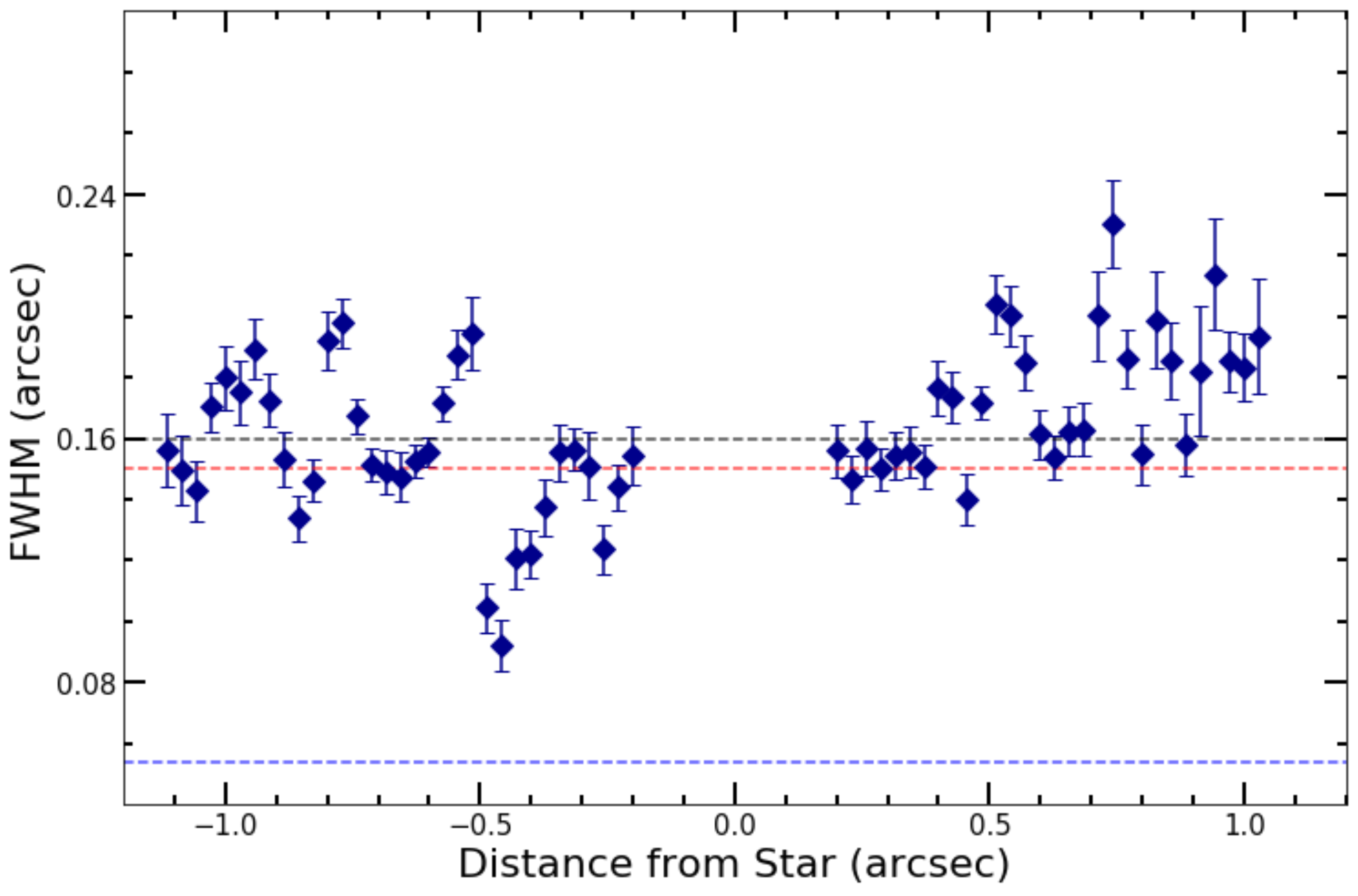}
\caption{\label{fig:fig3}$H$-band FWHM profile of the disk as a function of stellocentric distance in arcseconds. The weighted mean FWHM of the disk is represented by the grey dashed line, the intrinsic FWHM is represented by the red dashed line and the instrumental FWHM is represented by the blue dashed line.}
\end{figure}

To quantify the structure of the disk, we first measured the vertical height via the full width at half maximum (FWHM) of the disk as a function of radial separation from the star. We chose to do this with the $H$-band as it has the highest S/N. We first rotate our image so that the disk's major axis is horizontal. Next, we measure the surface brightness along vertical cross-sections of the disk, at various radial separations from the star until the whole disk has been covered. We then fit a Gaussian to each vertical brightness profile and use this to extract the FWHM at each cross-section. The results can be seen in Figure~\ref{fig:fig3}, where error bars are taken from fitting the Gaussian to the data. 

On both sides we measure the FWHM out to $r\approx1.1''$. We find the weighted mean FWHM (weights = 1/$\sigma^{2}$) to be 0.16$''$, represented by the grey dashed line in Figure~\ref{fig:fig3}. Comparing this value to the instrumental FWHM of the PSF ($\sim$0.054$''$), confirms that the disk is fully resolved. This value of 0.16$''$ is also $\sim$18\% larger than the value measured in Kalas et al. (\citeyear{Kalas:2015aa}) of $\sim$0.13$''$, although still consistent within the scatter of our FWHM data points. Subtracting the instrumental FWHM from our weighted mean in quadrature, gives us an intrinsic FWHM of $\sim$0.15$''$, or 15.6 AU. While the FWHM fluctuates around the weighted mean, there appear to be no significant trends in the FWHM with stellocentric distance. This is consistent with a radially narrow ring, as projection effects would cause an increase in the vertical FWHM with distance if the disk were broad. A narrow ring also aligns with the conclusions made by Kalas et al. (\citeyear{Kalas:2015aa}) and Lagrange et al. (\citeyear{Lagrange:2016aa}), however, we note that the vertical FWHM is only an indirect measure of the radial width and we cannot make a firm conclusion based on it alone. 

\subsection{Vertical Offset} \label{sec:vert_offset}
\begin{figure}
\includegraphics[scale=.32]{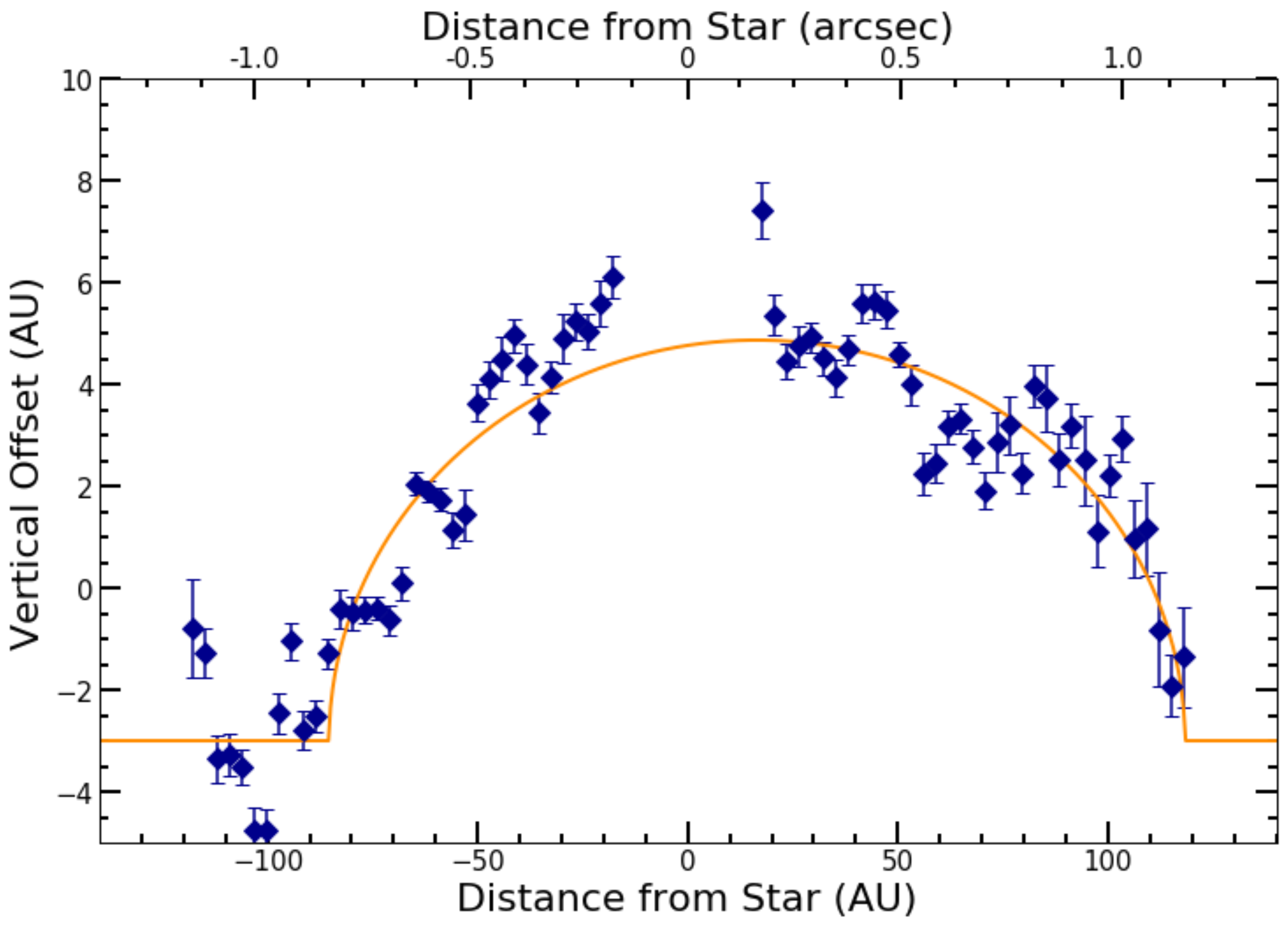}
\caption{\label{fig:spine}Vertical offset between the spine of HD 106906's disk and the star, as a function of radial separation shown in AU and arcseconds. The orange line represents the inclined ring model that best fits the spine.}
\end{figure}

While the vertical height tells us about the disk structure, measuring the vertical offset, which traces the disk spine, may be even more revealing. The disk spine refers to where the peak surface brightness, along the disk, is located relative to the star in the vertical direction. To measure this, we refer back to our Gaussian fitting (Section~\ref{sec:fwhm}), where the spine location is related to the mean value of the Gaussian. We can then compare the vertical offset between the spine location and the star as a function of radial separation. These results can be seen as the blue data points in Figure~\ref{fig:spine}, where the error bars are taken from the Gaussian fitting. At first glance, the vertical offset looks to be asymmetric, where past 100 AU ($\sim$1.0$''$) the vertical offset dips below -2 AU in the SE-extension, whereas this is not the case for the NW-extension.

To investigate this asymmetry further, we generate a simple inclined, circular, narrow ring model, following the same procedure used by Duch\^{e}ne et al. (\citeyear{Duchene_2020}). Our free parameters include the ring radius ($R_{d}$), inclination ($i$), position angle (PA) measured from East of North, and offsets of the ring center from the star along the major-axis and minor-axis ($\delta_{x}$ and $\delta_{y}$). Here, we are assuming that the star is located perfectly at the center of our data, however, it is important to note that some small systematic uncertainties may be present. A best fitting model is then found by using a Markov Chain Monte Carlo (MCMC) fit via the Python package $emcee$ (\citealt{Foreman-Mackey:2013aa}), for more details about this method see Section~\ref{sec:level5}. Our best fitting model can be seen in Figure~\ref{fig:spine} represented by the orange line. Through this modelling we find $R_{d}=101.84^{+1.75}_{-1.61}$ AU, $i=85.57^{+0.14}_{-0.15}$ degrees, and PA $=104.31^{+0.16}_{-0.30}$ degrees, where our inclination and PA are consistent with those derived in Kalas et al. (\citeyear{Kalas:2015aa}) and Lagrange et al. (\citeyear{Lagrange:2016aa}). More interestingly, we find $\delta_{x}=16.48^{+2.04}_{-1.76}$ AU, and $\delta_{y}=-2.99^{+0.28}_{-0.29}$ AU, which implies a significant eccentricity, placing the star closer to the SE-extension. While possible systematic uncertainties may lower the significance of the small offset along the minor-axis, this is not the case for the large offset along the major-axis. Therefore, we can use $\delta_{x}$ to place a lower limit on an eccentricity of $\gtrsim$0.16. Because modelling the spine suggests an inherently asymmetrical disk geometry, a circular ring model is no longer appropriate, where modelling with a proper ellipse will be needed to further probe the eccentricity and disk orientation.

To evaluate whether or not the x-offset is reasonable, we estimate the expected surface brightness asymmetry from a 16.48 AU offset by comparing $1/r^{2}$ (where $r$ is the distance from the star to disk) and the SPF between the SE and NW at the same radial separations from the star. We first calculate $r$ on both sides between 0.45$''$-0.50$''$, roughly half way across the disk, where the SPF would not be affected by limb brightening near the ansae or the noise present close to the star. Using a disk radius of 101.84 AU and a stellar offset of 16.48 AU towards the SE-extension, we obtain a SE/NW $1/r^{2}$ ratio of $\sim$1.4. This means that based solely on the relative distances between the star and the disk, we would expect the SE-extension to be about 1.4 times brighter than the NW-extension at those separations, about 12\% greater than what we measure with our $H$-band data (see Table~\ref{tab:asymm}). However, the difference between the SPF at those separations can also affect the expected brightness asymmetry. We therefore calculate the scattering angles between 0.45$''$ and 0.50$''$ and assume that the SPF in the HD 106906 disk conforms with the generic SPF observed in the Solar System and other debris disks (\citealt{Hughes:2018aa}) to estimate the SPF ratio between the SE and NW. Doing so, using our calculated scattering angles of $\sim$25-27 degrees for the NW and $\sim$29-33 degrees for the SE, we find that the NW-extension has a SPF $\sim$1.25 times greater than the SE-extension. Taking this into account, we expect that the greater SPF in the NW-extension partially cancels out the $1/r^{2}$ effect and therefore decreases the expected brightness asymmetry. While these are only rough estimates, as we do not know the exact SPF of this disk, they show that we would not expect a large brightness asymmetry despite the large offset derived from our spine fitting and that the more moderate brightness asymmetry observed with our data is reasonable.

\section{MCFOST Modelling of HD 106906} \label{sec:level4}

\subsection{Debris Disk Model} \label{sec:level4.1}
MCFOST is a 3-D radiative transfer code, designed to model circumstellar disks around young objects and can create fully polarized, scattered light, synthetic observations (\citealt{Pinte:aa}). For our models, the “debris disk” option is chosen within MCFOST, where the volume profile is defined by the following equation (\citealt{Aug99}):

\begin{eqnarray}
\rho(r,z)\ \propto \ \frac{\exp\Big( -\Big(\frac{\left| z \right|}{h(r)}\Big)^{\gamma}\Big)}{\Big[\Big(\frac{r}{Rc}\Big)^{- 2\alpha_{\text{in}}} + \Big(\frac{r}{Rc}\Big)^{- 2\alpha_{\text{out}}}\Big]^{1/2}}
\label{eq:one}.
\end{eqnarray}

\begin{eqnarray}
h(r)\ = \ H_{0} \Big(\frac{r}{R_{0}} \Big)
\label{eq:two}
\end{eqnarray}

In Equation~(\ref{eq:one}), \emph{r} is the radial distance from the star in the plane of the disk and $R_{c}$ is the critical radius inside and outside of which the surface density approaches power laws with indices $\alpha_{in}$ and $\alpha_{out}$, respectively. In addition, \emph{z} is the height above the disk mid plane and \emph{h(r)} describes the scale height profile, with a vertical exponent of $\gamma$. Equation~(\ref{eq:two}) defines \emph{h(r)}, where $H_{0}$ is the scale height at reference radius, $R_{0}$.

To model the dust grain properties Mie theory is used, where the dust grains are assumed to be spherical and compact. This idealized model, which does not capture the complexities expected in real dust grains, has implications that will be discussed later in this paper. However, for now, this assumption allows for easy and fast scattering computations. In addition, the dust grain size follows a power law distribution of $dN(a) \propto a^{-q} da$, where $a$ ranges from the minimum dust grain size $a_{\text{min}}$ up to a size of 1 mm, and \emph{q} is the power law for the dust grain size distribution. We also explore dust composition, looking exclusively at the volume fractions of astro-silicates, amorphous carbon and water ice. The optical indices for each material are obtained from literature (Si, \citealt{draine84}; aC, \citealt{rouleau91}; H$_{2}$O, \citealt{Li98}). The dust compositions (Si, aC, and H$_{2}$O) are set using effective medium theory, in that they make up one dust grain species (Ex., a single grain could consist of 50\% Si, 25\% aC, and 25\% H$_{2}$O). First, the porosity is fixed for all dust grains which assume a single power law size distribution. Then, the composition of the dust grains is allowed to vary as the relative fractions of each component are kept as free parameters. 

MCFOST creates each model at a user-specified wavelength. For our purposes, we choose 1.647 $\mu$m (where the $H$ filter throughput peaks). MCFOST produces a 4-dimensional fits image, which includes the final model in Stokes I, Q, U, and V intensities. The Q and U maps for each model are converted to Stokes Q$_{\phi}$, and are convolved with an instrumental PSF. Both our $H$-band data and models are masked at radial separations greater than 1.2$''$ and smaller than 0.12$''$ to ensure that noisy pixels are left out.

\subsection{MCMC Analysis} \label{sec:level5}

To determine the best-fitting values for each parameter explored, the models must be compared to the data. To do this, we again utilize the Python package \textit{emcee} (\citealt{Foreman-Mackey:2013aa}). Emcee uses MCMC, which allows us to find confidence intervals and best estimates for each parameter defining the density structure and dust grain properties of the disk. 

In this process, we model the $H$-band data alone, as it has the highest S/N. To begin with, properties that are related to the density structure of the disk are first allowed to vary over a range of priors. The parameters that would most strongly affect the disk brightness are kept constant; this includes dust grain properties such as the minimum grain size and composition. This is done to limit parameter space, and therefore speed up computation time. A list of density structural parameters, the initial value for each, and their prior range, can be viewed in Table~\ref{tab:table2}.

\begin{table}
\caption{\label{tab:table2}%
The density structural parameters and dust grain properties chosen to be constrained, along with the initial value and prior limits chosen for MCMC analysis.
}
\begin{ruledtabular}
\begin{tabular}{c c c}
\multicolumn{1}{c}{\textrm{Param.}}&\textrm{Initial Val.}&\textrm{Limits}\\
\colrule
H$_{0}$ [AU] & 5.0 & [1, 10] \\
$\gamma$ & 2.0 & [0.1, 3] \\
$\text{R}_{\text{c}}$ [AU] & 100 & [50, 200] \\
$\alpha_{\text{in}}$ & 1.0 & [0.5, 15] \\
$\alpha_{\text{out}}$ & -1.0 & [-5, -0.5] \\
x$_{*}$ [AU] & 0.0 & [-40, 40] \\
z$_{*}$ [AU] & 0.0 & [-10, 10] \\
M$_{\text{dust}}$ [M$_{\oplus}$] & 0.033 & [0.00033, 3.33] \\
$\text{porosity}$ & 0.5 & [0, 1] \\
$\text{V}_{\text{Si}}$ & 0.5 & [0, 1] \\
$\text{V}_{\text{aC}}$ & 0.5 & [0, 1] \\
$\text{V}_{\text{H}_{2}\text{O}}$ & 0.5 & [0, 1] \\
$\text{a}_{\text{min}}$ [$\mu$m] & 1.5 & [0.3, 4] \\
q & 3.5 & [2, 6] \\
\end{tabular}
\end{ruledtabular}
\end{table}

Our free parameters for the density structure include $H_{0}$, $\gamma$, $R_{\text{c}}$, $\alpha_{in}$, and $\alpha_{out}$. We also include stellar offset parameters, $x_{*}$ and $z_{*}$, where $x_{*}$ is the offset along the disk major-axis and $z_{*}$ is the offset in the vertical direction relative to the disk. The parameter, $x_{*}$, is related to $\delta_{x}$ from our disk spine fitting in Section~\ref{sec:vert_offset}, where the main difference is that the star itself is being offset instead of the inclined ring, meaning that an offset towards the SE-extension would be negative rather than positive. The parameter, $z_{*}$, differs from $\delta_{y}$ as it represents an offset above or below the disk mid-plane, where $\delta_{y}$ is the projected offset along the minor-axis. While our parameterized models do not account for geometrical asymmetries (the synthetic disk remains circular), as to not make our models overly complex, a stellar offset can still be used to create a surface brightness asymmetry. Important structural parameters such as inclination and PA were not considered in this analysis, as they are already well defined from previous work, as well as our spine fitting, to be $\sim$85 degrees and $\sim$104 degrees, respectively (\citealt{Lagrange:2016aa}; \citealt{Kalas:2015aa}). Therefore, we fix these parameters in order to reduce parameter space and computational time. Additionally, the inner radius is set to 50 AU, based on observations from Kalas et al. (\citeyear{Kalas:2015aa}), and the outer radius to 200 AU (roughly twice the outer radius observed in our data), where $R_{0}$ is fixed at 100 AU.

After we model the density structure, we then hold these values constant at the median (50th) percentile values and instead vary the dust grain parameters. A list of the dust grain parameters, the initial value for each, and their prior range can be viewed in Table~\ref{tab:table2}. Here, $M_{dust}$ is the total dust mass in Earth masses, porosity defines how fluffy or compact the dust grains are (where zero porosity corresponds to a perfectly compact dust grain), $V_{Si}$ is the volume fraction of astronomical silicate, $V_{aC}$ is the volume fraction of amorphous carbon, and $V_{H_{2}O}$ is the volume fraction of water ice. Finally, $a_{min}$ is the minimum dust grain size, and $q$ is the power law for the distribution of dust grain sizes.

For each run, 200 walkers are deployed to explore the parameter space, starting at the initial values as stated in Table~\ref{tab:table2}. These walkers move around freely in order to find the maximum log likelihood values for each parameter, as well as their posterior distributions. The likelihoods between our models and $H$-band data are calculated pixel-by-pixel using the $\chi^{2}$ function. This occurs for up to several thousand iterations until after the parameters have clearly converged, or the log likelihood for the walkers is no longer increasing.

\subsection{Results} \label{sec:level6}

\begin{figure*}
\centering
\includegraphics[scale=.44]{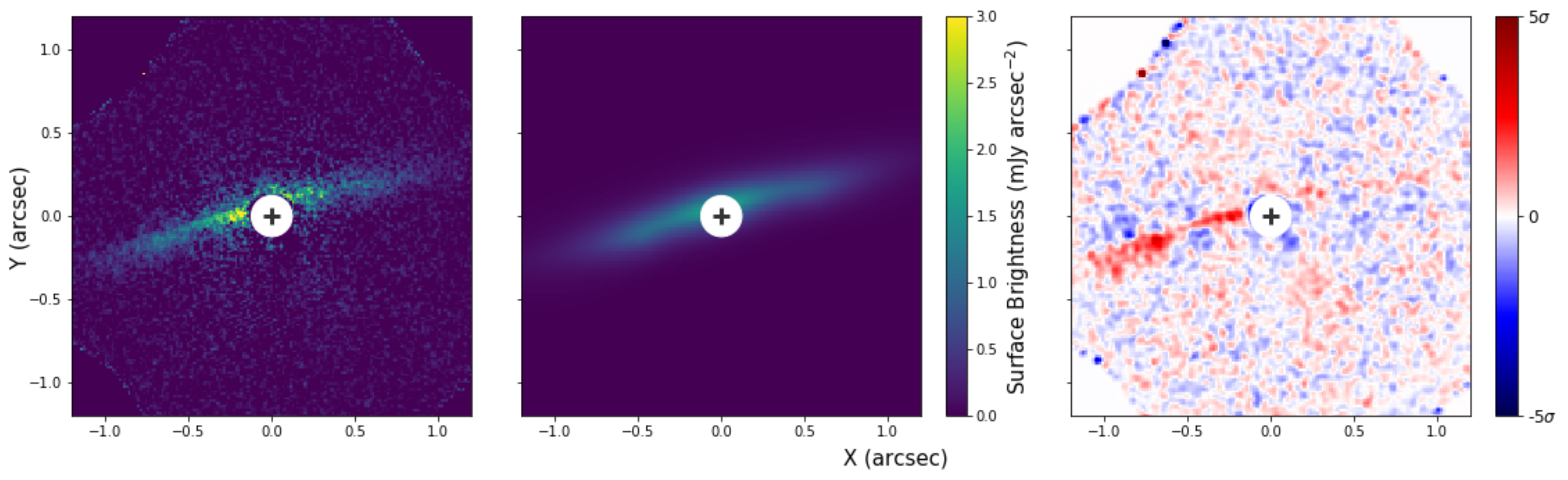}  
\caption{\label{fig:fig5}Results from modelling the whole disk with $H$-band only. \textbf{Left:} $H$-band data. \textbf{Middle:} Final median model. \textbf{Right:} Residual between data and final median model.}
\end{figure*}

For the MCMC analysis of the density structural parameters, emcee was allowed to run for a few thousand iterations, until well after convergence to make sure that we had the best results. We then reran the MCMC analysis, where this time the density structural parameters were kept fixed using the median percentile values (50\%) obtained in the previous run. This analysis was allowed to run again until the log likelihood was no longer noticeably changing. The results for both the density structure and dust grain parameters can be found in Table~\ref{tab:table4} which lists the 16th, 50th, and 84th percentiles. The final model, created from the median percentile values from each parameter's confidence intervals, can be seen in Figure~\ref{fig:fig5}. 

From our MCMC modelling, we obtain an $H_{0}$ of $4.34^{+0.04}_{-0.03}$ AU with a vertical profile exponent of $\gamma=0.715^{+0.01}_{-0.01}$. This would give the scale height to disk radius a ratio of $h/r \approx 4\%$, which is consistent with $h/r$ found for other debris disks such as $\beta$ Pic, Fomalhaut, and AU Mic (\citealt{Millar-Blanchaer:2015aa}; \citealt{Kalas:2005aa}; \citealt{Krist_2005}). However, the shallow $\gamma$ appears to be in contradiction to the Gaussian profile used in Section~\ref{sec:fwhm}, driven by the shape of our data, where a shallow $\gamma$ would suggest a more Lorentzian profile. Given our limited S/N and model limitations, we note it may be difficult to firmly establish the nature of the vertical profile. Next we obtain a critical radius, $R_{c}$, of $72.21^{+3.10}_{-3.30}$ AU, which lies roughly half way between the set inner radius (50 AU) and our disk radius of $101.84$ AU derived in Section~\ref{sec:vert_offset}. We also find surface density power laws of $\alpha_{in}=1.03^{+0.29}_{-0.24}$ and $\alpha_{out}=-2.26^{+0.11}_{-0.08}$. These shallow power laws give the disk a dust density distribution ranging from a fixed inner radius of 50 AU to a half-maximum at $\sim$111 AU and a $R_{peak}$ (radius of peak density) of $\sim$64 AU, consistent with that derived in Lagrange et al. (\citeyear{Lagrange:2016aa}). However, this is in contradiction to the assumption of a radially narrow ring based on our vertical FWHM profile, as these values give $\Delta$R/R=0.95, consistent with a radially broad disk. Although projection effects may still be present within our limited S/N data, another possibility is that the inner radius lies beyond 50 AU. Given that the near-edge configuration of the disk makes it difficult to constrain $R_{in}$, we can only put an upper limit on $\Delta$R/R of 0.95. Future modelling and analysis of polarized and spectral data may be able to help shed more light into the location of the inner radius. 

For the stellar offset parameter, $z_{*}$, we find values that are very close to 0 (0.25-0.29 AU), meaning that the star lies within the mid-plane of the disk as we would expect. Surprisingly, the MCMC favors a stellar offset of only $\sim$3.58 AU towards the SE-extension of the disk, which is much smaller than the stellar offset of $\sim$16.5 AU predicted from our spine fitting. An offset this small is unable to match the surface brightness asymmetry observed in the $H$-band. This can be seen in Figure~\ref{fig:fig5}, where there is a large residual remaining in the SE-extension. One reason why the MCMC might prefer models with a smaller stellocentric offset could be the lack of a radial extent asymmetry in our data, as a larger offset would create a truncated SE-extension. This would be an issue caused by the fact that, despite having a stellocentric offset, our models are geometrically symmetric circles.
 
\begin{table}[t]
\caption{\label{tab:table4}%
MCMC results for density structural and dust grain parameters. This includes the values for the 16th, 50th, and 84th percentiles.
}
\begin{ruledtabular}
\begin{tabular}{c c c c}
\multicolumn{1}{c}{\textrm{Param.}}&\textrm{16}$\%$&\textrm{50}$\%$&\textrm{84}$\%$\\
\colrule
H$_{0}$ [AU] & 4.07 & 4.34 & 4.38 \\
$\gamma$ & 0.708 & 0.715 & 0.723 \\
$\text{R}_{\text{c}}$ [AU] & 68.92 & 72.21 & 75.31 \\
$\alpha_{\text{in}}$ & 0.79 & 1.03 & 1.32 \\
$\alpha_{\text{out}}$ & -2.34 & -2.26 & -2.16 \\
x$_{*}$ [AU] & -3.63 & -3.58 & -3.35 \\
z$_{*}$ [AU] & 0.25 & 0.27 & 0.29 \\
M$_{\text{dust}}$ [M$_{\oplus}$] & 0.012 & 0.02 & 0.11 \\
$\text{porosity}$ & 0.02 & 0.1 & 0.29 \\
$\text{V}_{\text{Si}}$ & 0.19 & 0.39 & 0.53 \\
$\text{V}_{\text{aC}}$ & 0.06 & 0.38 & 0.70 \\
$\text{V}_{\text{H}_{2}\text{O}}$ & 0.04 & 0.20 & 0.46 \\
$\text{a}_{\text{min}}$ [$\mu$m] & 0.91 & 3.17 & 3.72 \\
q & 2.99 & 3.19 & 3.30 \\
\end{tabular}
\end{ruledtabular}
\end{table}

While our confidence intervals for the density structure parameters were well constrained, constraining the dust grain parameters proved to be much more difficult. We obtain a total dust mass of $0.02^{+0.09}_{-0.008}$ M$_{\oplus}$. This is on the same order as the masses derived from millimeter (${\sim}0.05$ M$_{\oplus}$; \citealt{Kral:2020aa}) and mid-IR observations (0.067 M$_{\oplus}$; \citealt{Chen2011}); however, our confidence interval is large and spans two probability peaks at ${\sim}$0.016 M$_{\oplus}$ and ${\sim}$0.11 M$_{\oplus}$. We obtain a porosity of $0.10^{+0.19}_{-0.08}$, showing a significant preference for grains with a low porosity or compact grains. The composition is nearly unconstrained, with all three compositions (Si, aC, and H$_{2}$O) having very large confidence intervals and multiple peaks in their posterior distributions. The MCMC results slightly favor larger fractions of Si and aC, with smaller fractions of H$_{2}$O; however, we cannot make strong conclusions about the composition given their broad posteriors. 

For the minimum dust grain size we obtain $a_{min}=3.17^{+0.55}_{-2.26}$ $\mu$m, where we have a large probability peak at 0.91 $\mu$m and a second, smaller peak between 3-4 $\mu$m. While a minimum dust grain size of 0.91 $\mu$m is more consistent with the calculated blowout size for the system ($\sim$0.85 $\mu$m)\footnote{Assuming a combined mass of 2.71 M$_{\odot}$ for HD 106906AB (\citealt{Rodet:2017aa}), and a luminosity of 6.56$\pm$0.04 L$_{\odot}$ (\citealt{Collaboration:2018aa}), we approximate the blowout size (a$_{\text{blow}}$) for this system to be $\sim$0.85 $\mu$m (\citealt{Pawellek:2015aa}; \citealt{Burns79}). Dust grains below this size should be blown out of the disk by radiation pressure from the central binary.}, a larger size may be more consistent with the disk color measured in Section~\ref{sec:flux}. To test this, we compute $\Delta$(M$_{\text{disk}}$ arcsec$^{-2}$ - M$_{\text{star}}$), over the same integrated distances as in Section~\ref{sec:flux}, for our median model at all three wavelengths. This is done for both an $a_{min}$ of 0.91 $\mu$m and 3.17 $\mu$m. We find that an $a_{min}$ of 3.17 $\mu$m is more consistent with the observed disk color ($H$-$K1$=-0.15 mag), compared to our model with an $a_{min}$ of 0.91 $\mu$m, which exhibits a much stronger blue color ($H$-$K1$=-0.66 mag). Therefore, we find the data favor a larger minimum dust grain size, although, $a_{min}$ could be further informed in future work using the degree of linear polarization. 

Finally, we obtain a power law for the dust grain size distribution of $q=3.19^{+0.11}_{-.20}$. This is slightly smaller than the predicted size distribution slope of $q\approx 3.5$ for a collisional cascade (\citealt{Dohnanyi69}; \citealt{Marshall:2017aa}), however, other debris disks have been found to deviate from this value. Additionally, our obtained value for $q$ is comparable to the size distribution slope found for the mm-observations of $3.36\pm0.13$ (\citealt{Kral:2020aa}) within 1$\sigma$.

Comparing our $H$-band data with the best-fit model, we see that while the NW-extension seems to be modelled well with no lingering residuals, as stated before, the SE-extension features a significant residual along the disk. We find this to be more evidence for a geometrically asymmetric and likely eccentric disk, as our symmetric/circular models are unable to fully model both the disk structure and surface brightness asymmetry simultaneously. This may have affected our final modelling results, since we would need a model that has an asymmetric dust density profile to more accurately model our asymmetric disk. However, we leave this more complex modelling for future work.

\section{Discussion} \label{sec:level7}

\subsection{Model Limitations} \label{sec:limits}
There are several limitations with our MCFOST models that could impact our results. The first, as discussed above in Section~\ref{sec:level6}, is that our models are not inherently eccentric. While we are able to offset the star, we assume the disk to be circular with a symmetric dust density profile, which would not be the case for a truly eccentric disk. This assumption has likely caused our final model to underestimate the surface brightness asymmetry as seen in the residual in Figure~\ref{fig:fig5}. Following the same steps as in Section~\ref{sec:vert_offset}, we compute $1/r^{2}$ and the SPF between 0.45$''$ and 0.50$''$, this time using a stellar offset of 3.58 AU. Doing so, we find that the effects from $1/r^{2}$ and the SPF at these radial separations almost completely cancel each other out, meaning that no brightness asymmetry would be seen with an offset this small. While a larger offset is needed to reach the observed brightness asymmetry, increasing the stellar offset in our symmetric models creates a more ill-fitting model as this creates a radial extent asymmetry not observed in our data. This leads us to believe that in order to better represent the disk, we need a model that is eccentric with an asymmetric dust density profile.

Another limitation, is that our models assume a smooth dust mass distribution defined with only one power-law, which may not be necessarily realistic, as certain sizes of dust grains may be missing or the distribution may more closely resemble a function composed of more than one power-law (\citealt{Strubbe:2006aa}). High resolution ALMA data would be useful in this case to constrain the dust mass distribution with millimeter observations, which can then be used within MCFOST to create a more realistic model.

The final limitation comes from the use of Mie theory in our models. As mentioned before, Mie theory uses compact spherical grains, which makes computations much faster and easier, but is unrealistic. In reality, dust grains are not perfectly spherical, but are rather irregularly shaped aggregates. This may strongly affect the accuracy of our results for the dust grain properties, as they rely heavily on this assumption and how light is scattered. For example, Min et al. (\citeyear{Min:2015aa}) found the SPF differed between Mie dust grains and realistic aggregates for scattering angles greater than 90 degrees (backside of the disk). Where for Mie grains, the SPF generally decreased, they found that for aggregates, the SPF either increased or stayed flat. Another discrepancy due to Mie theory, is the affect this has on the blowout size. Arnold et al. (\citeyear{Arnold:2019aa}) calculated a$_{\text{blow}}$ for aggregates with porosity of 76.4$\%$, and found that, in general, it was 2-3 times higher than for Mie compact grains. In addition, multiple studies of other debris disks have also found that Mie theory is unable to match observations (\citealt{Milli_2017}; \citealt{Duchene_2020}; \citealt{Arriaga_2020}). While we leave more complex modelling with realistic aggregates for future work, we believe our results are still useful in comparing to previous work done on HD 106906 and other debris disks.

\subsection{Dynamical Interaction with HD 106906b}
In this section, we compare our results to previous simulation work, including both works focused specifically on HD 106906 and more general disk-planet interaction studies.

The empirical analysis of our data supports a significant eccentricity; this could be due to interaction with HD 106906b, either as it was ejected from the disk, or due to secular orbital effects. There are two previous studies that consider these scenarios using particle simulations (\citealt{Nesvold:2017aa}; \citealt{Rodet:2017aa}). While Rodet et al. (\citeyear{Rodet:2017aa}) examines the morphological outcomes of the disk from both the planet being ejected and from secular effects, Nesvold et al. (\citeyear{Nesvold:2017aa}) examines solely the secular effects.

Rodet et al. (\citeyear{Rodet:2017aa}) found that the ejection of the planet from the disk was able to reproduce the asymmetries on larger scales as seen with HST, where particles are blown out to high eccentricities on the right side of the disk. However, the ejection was not able to reproduce the SE/NW asymmetry on smaller scales as seen with GPI in this work and Kalas et al. (\citeyear{Kalas:2015aa}). On the other hand, the effect of the planet on a eccentric, inclined orbit, was able to reproduce the smaller scale asymmetries with a larger density in the SE than the NW. This also gives rise to a needle like halo at larger distances, but not as strongly as the ejection scenario, suggesting that the two effects together could explain the observations on large and small scales.

In contrast, Nesvold et al. (\citeyear{Nesvold:2017aa}) conducted an in-depth study of the effect on the disk from the planet, initially located at 700 AU (in situ) with an adopted eccentricity of 0.7 as it orbits at different inclinations. Through this alone, they were able to recreate both the small and large scale asymmetries, where a needle-like structure was seen up to 500 AU, while closer in the SE-extension is brighter than the NW-extension. The simulation also predicts that the planet should induce a large eccentricity on the disk, up to $\sim$0.3, which agrees well with the large eccentricity (e$\gtrsim$0.16) we derive from our spine fitting. One caveat is that this model predicts a greater surface brightness asymmetry than we observe in our GPI data, particularly past 50 AU ($\sim$50\% brighter SE-extension compared to $\sim$18-27\% derived in Section~\ref{sec:flux}), however, this is taking into account the total scattered light compared to polarized light.

In a more general case, Lee \& Chiang (\citeyear{Lee:2016aa}) look at a number of different debris disk morphologies induced by a single, eccentric planet companion. One of these morphologies is the ``Needle", which is described as an eccentric and vertically thin disk viewed nearly edge on. In their analysis, they find that a single planet with eccentricity of 0.7 creates a radial length asymmetry, where one side extends much farther than the other (the needle), as well as a brightness asymmetry. In this case, the shorter side of the disk is brighter closest to the star, but eventually at larger distances, the longer side becomes brighter. This is consistent with observations of the HD 106906 debris disk, as the ``Needle" is observed with HST at large radial separations in the NW-extension, where a brightness asymmetry is observed with GPI at smaller radial separations. While in these simulations, the planet has a mass of 10 M$_{\oplus}$ and is orbiting interior to the disk, the results produced are similar to the simulations done with a larger planet on an eccentric orbit outside of the disk for HD 106906. While we cannot fully discriminate against a planet perturber interior to the disk, HD 106906b is still a likely cause for the observed disk morphology. 

These simulations show that a single, eccentric planet on both an internal and an external orbit can cause a debris disk to become eccentric, as well as produce a brightness asymmetry consistent with what we observe with our GPI data. Additionally, Nesvold et al. (\citeyear{Nesvold:2017aa}) and Lee \& Chiang (\citeyear{Lee:2016aa}) show that this is sufficient to create the needle like halo that is observed with HST. The recent constraints of HD 106906b's orbit also indicates that it most likely has a misaligned and eccentric orbit, with a periastron distance that comes reasonably close to the star (\citealt{Nguyen2020}). This is similar to the conditions used for the planet's orbit in Nesvold et al. (\citeyear{Nesvold:2017aa}) which are able to reproduce both the small and large scale asymmetries, providing further evidence for the planet being the disk's perturber. Based on simulation work and our observations, dynamical interaction with HD 106906b remains a strong candidate for the source of the disk's asymmetries.

\subsection{ISM Interaction}
Another possibility for the observed surface brightness asymmetry is an interaction with the ISM (\citealt{Debes:2009aa}). If a debris disk passes through a dense cloud of interstellar gas, small grains can be stripped from the disk into the halo due to ram pressure. This has been used as one attempt to explain the morphological features in other debris disks, such as the "wing" feature for HD 32297 and HD 61005 (\citealt{Schneider:2014aa}), as well as the "needle" feature for HD 15115 (\citealt{Rodigas:2012aa}).

An ISM interaction comes up as a scenario for HD 106906 because of the disk's asymmetrical needle-like halo, where passing through a dense ISM cloud would cause small dust grains to be blown into the NW-extension of the halo as observed in the optical with HST (\citealt{Kalas:2015aa}). For this to occur, we would expect the proper motion of the binary to be in the opposite direction of the needle (SE-direction). HD 106906AB has a proper motion of -39.014 mas/yr in RA and -12.872 mas/yr in Dec. (\citealt{Collaboration:2018aa}), meaning that the proper motion is pointing in the SW-direction. This is inconsistent with what we would expect for an ISM interaction. In addition, we do not see a difference in the color along the disk, where we might expect an ISM interaction to cause the NW-extension of the disk to be more blue due to the direction of small grains blown and distributed towards the NW.

Taking these inconsistencies into consideration, we determine that an ISM interaction is unlikely to be the source of the disk's asymmetry. Based on our findings of an asymmetric structure and consistent disk color for both extensions, the source of asymmetry should be an event that changes the dust density distribution without greatly changing the dust grain size distribution. In this case, we find that dynamical interaction with the planet companion is more consistent with observations.

\section{\label{sec:level8}Conclusion}
With GPI, we have obtained deeper polarimetric data of HD 106906's disk in $H$-band, as well as $J$ and $K1$ bands. With these data, we conclude the following:

\begin{enumerate}
    \item We detect a surface brightness asymmetry along the disk's major axis in all three bands, where the SE-extension is brighter than the NW-extension. This asymmetry peaks within 0.4$''$ and again between 0.6$''$-0.8$''$ for the $H$-band. We also estimate the disk color by comparing the magnitude of the disk with the stellar magnitude in each band, where we find that the disk to have a blue color in polarized light. We also do not observe a difference in color between the two extensions, implying that the dust grain properties do not significantly vary across the disk. 
    \item For the structure of the disk, we find an intrinsic vertical FWHM of $\sim$0.15$''$ or 15.6 AU. While we do not observe a strong asymmetry in radial extent or in the vertical FWHM, we observe a strong asymmetry in the vertical offset. Fitting a simple inclined ring model, we find a disk radius of $\sim$101.9 AU, an inclination of $\sim$85.6$^{\circ}$ and a PA of $\sim$104.3$^{\circ}$, where the inclination and PA are consistent with those derived in Kalas et al. (\citeyear{Kalas:2015aa}) and Lagrange et al. (\citeyear{Lagrange:2016aa}). Additionally, we derive a disk offset of $\sim$16.5 AU along the major-axis and $\sim$3.0 AU along the minor-axis, both towards the SE-extension. This would give the disk a significant eccentricity of $\gtrsim$0.16, which could help to explain the surface brightness asymmetry observed. While an elliptical fitting is needed to further probe the eccentricity and disk orientation, this confirms that the disk is in fact geometrically asymmetric.
    \item Using MCFOST within an MCMC framework, we were able to model the disk and constrain its density structure and dust grain parameters. While our best model fits the NW-extension well, the surface brightness is underestimated in the SE-extension. We find this to be more evidence of the HD 106906 disk being structurally asymmetric, as our structurally symmetric models are unable to capture both the disk structure and surface brightness simultaneously.
    \item We considered two possible sources of the disk's asymmetry: Dynamical interaction with the planet companion, HD 106906b, and an interaction with the ISM. We find that our observations are more consistent with a disk shaped by interaction with the planet companion, where our analysis strongly suggests a disk that is asymmetrical in structure, but does not significantly vary in dust grain properties based on the disk color. An interaction with the ISM is deemed unlikely, as the proper motion of the system does not align with the direction we would expect in order to produce the needle-like halo.
\end{enumerate}

Our results reveal further insights into the structure and asymmetries of the HD 106906 debris disk. Compared to previous studies, we find similar trends in surface brightness as well as certain geometrical properties such as inclination and PA. Additionally, with our higher S/N $H$-band data and multi-wavelength $J$- and $K1$-band images, we have further constrained the structure and color of the disk. Most interestingly, our spine fitting reveals strong evidence of a disk that is geometrically asymmetric and likely eccentric, something that has not been shown before. For future work, we plan to apply an elliptical fitting to probe the eccentricity and disk orientation further. In the case of radiative transfer modelling, using a model that has an asymmetrical dust density distribution, as well as utilizing realistic aggregates, would help shed more light into the density structure and dust grain properties of the disk.

\newpage
\acknowledgments

The authors wish to thank the anonymous referee for helpful suggestions that improved this manuscript. This work is based on observations obtained at the Gemini Observatory, which is operated by the Association of Universities for Research in Astronomy, Inc. (AURA), under a cooperative agreement with the National Science Foundation (NSF) on behalf of the Gemini partnership: the NSF (United States), the National Research Council (Canada), CONICYT (Chile), Ministerio de Ciencia, Tecnolog\'{i}a e Innovaci\'{o}n Productiva (Argentina), and Minist\'{e}rio da Ci\^{e}ncia, Tecnologia e Inova\c{c}\~{a}o (Brazil). This work made use of data from the European Space Agency mission \emph{Gaia} (\url{https://www.cosmos.esa.int/gaia}), processed by the \emph{Gaia} Data Processing and Analysis Consortium (DPAC, \url{https://www.cosmos.esa.int/web/gaia/dpac/consortium}). Funding for the DPAC has been provided by national institutions, in particular the institutions participating in the Gaia Multilateral Agreement. This research made use of the SIMBAD and VizieR databases, operated at CDS, Strasbourg, France. We thank support from NSF AST-1518332, NASA NNX15AC89G and NNX15AD95G/NEXSS. This work benefited from NASA’s Nexus for Exoplanet System Science (NExSS) research coordination network sponsored by NASA's Science Mission Directorate. KC is supported by an NSERC Discovery Grant to BCM.

%

\vspace{5mm}
\facilities{Gemini:South}


\software{MCFOST (\citealt{Pinte:aa}),
Gemini Planet Imager Pipeline (\citealt{Perrin:2014aa}, \url{http://ascl.net/1411.018)},
emcee (\citealt{Foreman-Mackey:2013aa}, \url{http://ascl.net/1303.002}),
corner (\citealt{corner}, \url{http://ascl.net/1702.002}), 
matplotlib (\citealt{Hunter07}; \citealt{droettboom17}), 
iPython (\citealt{perez07}), 
Astropy (\citealt{Collaboration:2018ab}),
NumPy (\citealt{Oliphant_06}; \url{https://numpy.org}),
SciPy (\citealt{Virtanen_20}; \url{http://www.scipy.org/})}


\clearpage
\appendix

\section{\label{sec:level10}Corner Plots}
\begin{figure*}[h]
\centering
\includegraphics[scale=.44]{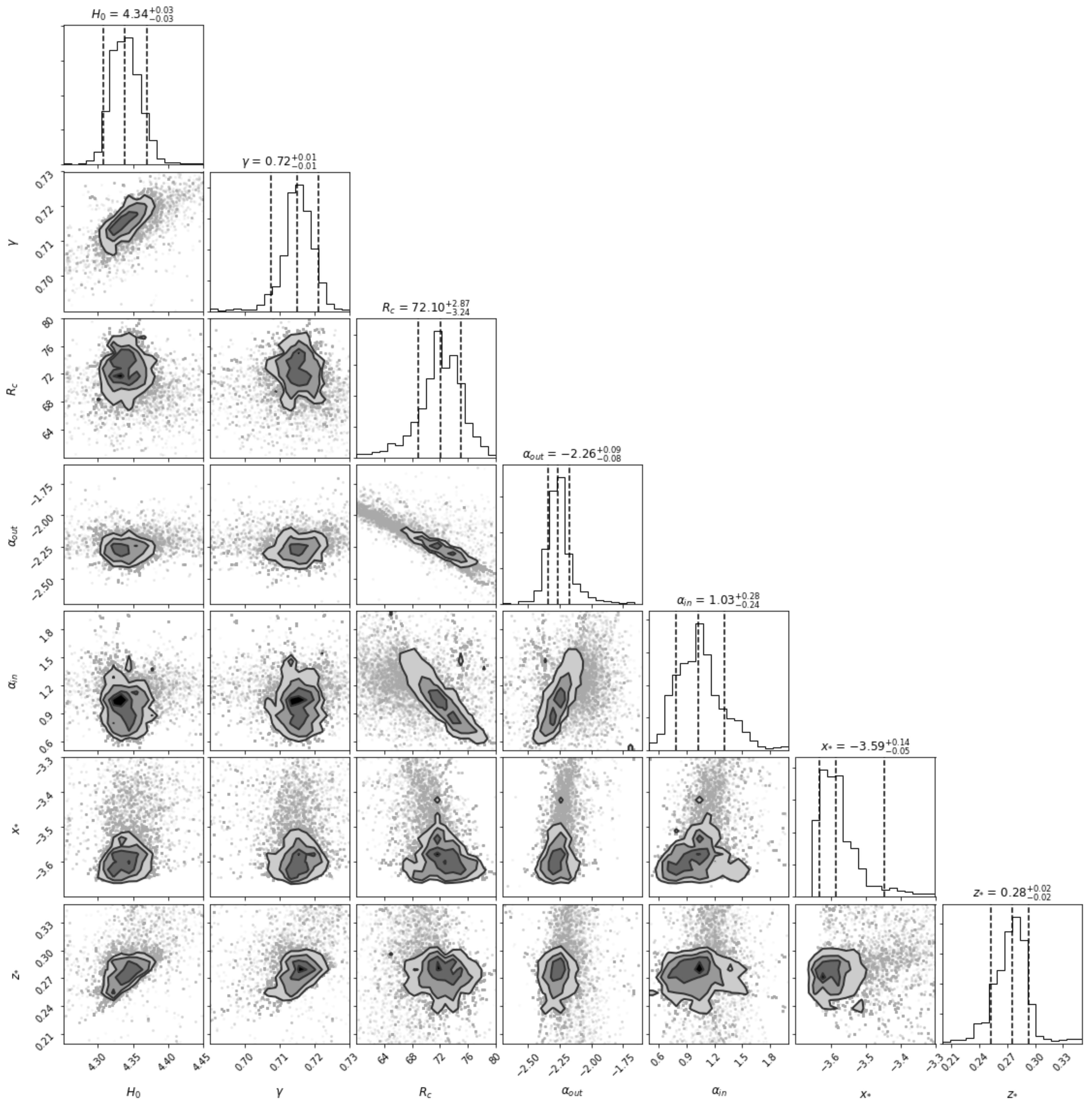}
\caption{\label{fig:fig10}Corner plot for MCMC results of density structural parameters.}
\end{figure*}

\begin{figure*}
\includegraphics[scale=.44]{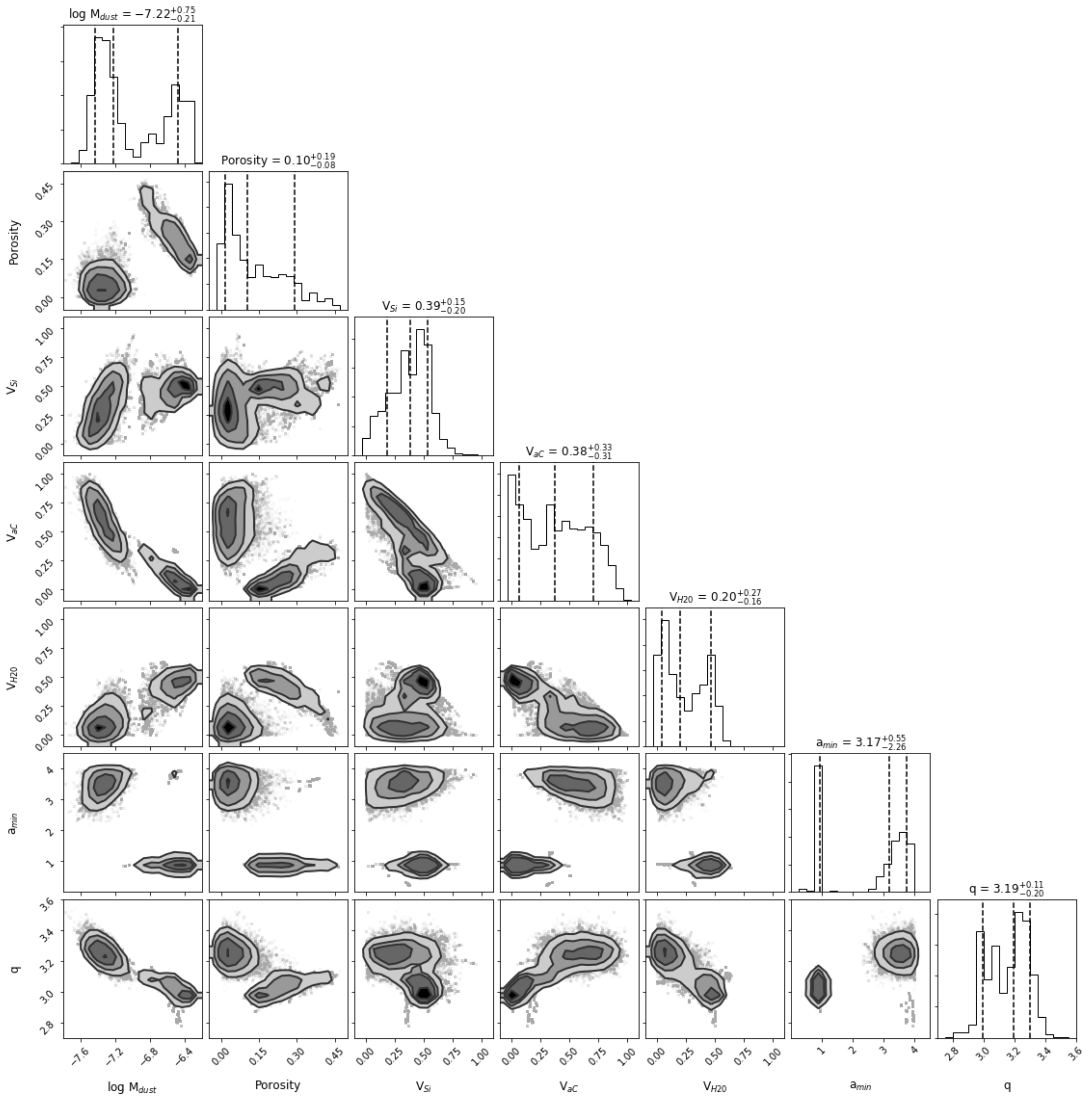}
\centering
\caption{\label{fig:fig11}Corner plot for MCMC results of dust grain parameters.}
\end{figure*}

Below are the corner plots from our modelling and MCMC analysis (\citealt{corner}). Figure~\ref{fig:fig10} shows the posterior distributions for the density structural parameters modelled with $H$-band only, while Figure~\ref{fig:fig11} shows the posterior distributions for the dust grain properties modelled with $H$-band only.

\clearpage


\bibliography{HD106906.bbl}

\begin{thebibliography}{}
\expandafter\ifx\csname natexlab\endcsname\relax\def\natexlab#1{#1}\fi
\providecommand{\url}[1]{\href{#1}{#1}}
\providecommand{\dodoi}[1]{doi:~\href{http://doi.org/#1}{\nolinkurl{#1}}}
\providecommand{\doeprint}[1]{\href{http://ascl.net/#1}{\nolinkurl{http://ascl.net/#1}}}
\providecommand{\doarXiv}[1]{\href{https://arxiv.org/abs/#1}{\nolinkurl{https://arxiv.org/abs/#1}}}

\bibitem[{Arnold {et~al.}(2019)Arnold, Weinberger, Videen, \&
  Zubko}]{Arnold:2019aa}
Arnold, J.~A., Weinberger, A.~J., Videen, G., \& Zubko, E.~S. 2019, ApJ, 157,
  157, \dodoi{10.3847/1538-3881/ab095e}

\bibitem[{Arriaga {et~al.}(2020)Arriaga, Fitzgerald, Duchêne, Kalas,
  Millar-Blanchaer, Perrin, Chen, Mazoyer, Ammons, Bailey, \&
  et~al.}]{Arriaga_2020}
Arriaga, P., Fitzgerald, M.~P., Duchêne, G., {et~al.} 2020, AJ, 160, 79,
  \dodoi{10.3847/1538-3881/ab91b1}

\bibitem[{Augereau \& Beust(2006)}]{Augereau:aa}
Augereau, J.~C., \& Beust, H. 2006, A\&A, 455, 987-999,
  \dodoi{10.1051/0004-6361:20054250}

\bibitem[{Augereau {et~al.}(1999)Augereau, Lagrange, Mouillet, Papaloizou, \&
  Grorod}]{Aug99}
Augereau, J.~C., Lagrange, A.-M., Mouillet, D., Papaloizou, J. C.~B., \&
  Grorod, P.~A. 1999, A$\&$A, 348, 557

\bibitem[{Bailey {et~al.}(2014)Bailey, Meshkat, Reiter, Morzinski, Males, Su,
  Hinz, Kenworthy, Stark, Mamajek, Briguglio, Close, Follette, Puglisi,
  Rodigas, Weinberger, \& Xompero}]{Bailey:2013aa}
Bailey, V., Meshkat, T., Reiter, M., {et~al.} 2014, ApJ, 780, L4,
  \dodoi{10.1088/2041-8205/780/1/L4}

\bibitem[{Boccaletti {et~al.}(2003)Boccaletti, Augereau, Marchis, \&
  Hanh}]{Boccaletti:2003aa}
Boccaletti, A., Augereau, J.~C., Marchis, F., \& Hanh, J. 2003, ApJ, 585,
  494-501, \dodoi{10.1086/346019}

\bibitem[{Burns {et~al.}(1979)Burns, Lamy, Soter, Lamy, \& Soter}]{Burns79}
Burns, J.~A., Lamy, P.~L., Soter, S., Lamy, P.~L., \& Soter, S. 1979, Icarus,
  40, 1

\bibitem[{Chen {et~al.}(2005)Chen, Jura, Gordon, \& Blaylock}]{chen05}
Chen, C., Jura, M., Gordon, K., \& Blaylock, M. 2005, ApJ, 623, 493,
  \dodoi{10.1086/428607}

\bibitem[{Chen {et~al.}(2011)Chen, Mamajek, Bitner, Pecaut, Su, \&
  Weinberger}]{Chen2011}
Chen, C.~H., Mamajek, E.~E., Bitner, M.~A., {et~al.} 2011, ApJ, 738, 122,
  \dodoi{10.1088/0004-637X/738/2/122}

\bibitem[{Daemgen {et~al.}(2017)Daemgen, Todorov, Quanz, Meyer, Mordasini,
  Marleau, \& Fortney}]{Daemgen:2017aa}
Daemgen, S., Todorov, K., Quanz, S.~P., {et~al.} 2017, A$\&$A, 608, A71,
  \dodoi{10.1051/0004-6361/201731527}

\bibitem[{{De Rosa} \& Kalas(2019)}]{Rosa:2019aa}
{De Rosa}, R.~J., \& Kalas, P. 2019, AJ, 157, 125,
  \dodoi{10.3847/1538-3881/ab0109}

\bibitem[{Debes {et~al.}(2009)Debes, Weinberger, \& Kuchner}]{Debes:2009aa}
Debes, J.~H., Weinberger, A.~J., \& Kuchner, M.~J. 2009, ApJ, 702, 318,
  \dodoi{10.1088/0004-637X/702/1/318}

\bibitem[{Dohnanyi(1969)}]{Dohnanyi69}
Dohnanyi, J. 1969, Journal of Geophysical Research, 74, 2531

\bibitem[{Draine \& Lee(1984)}]{draine84}
Draine, B.~T., \& Lee, H.~M. 1984, ApJ, 285, 89, \dodoi{10.1086/162480}

\bibitem[{Droettboom {et~al.}(2017)Droettboom, Caswell, Hunter, Firing, \&
  Nielsen}]{droettboom17}
Droettboom, M., Caswell, T.~A., Hunter, J., Firing, E., \& Nielsen, J.~H. 2017,
  Zenodo, \dodoi{10.5281/zenodo.573577}

\bibitem[{Duchêne {et~al.}(2020)Duchêne, Rice, Hom, Zalesky, Esposito,
  Millar-Blanchaer, Ren, Kalas, Fitzgerald, Arriaga, \& et~al.}]{Duchene_2020}
Duchêne, G., Rice, M., Hom, J., {et~al.} 2020, AJ, 159, 251,
  \dodoi{10.3847/1538-3881/ab8881}

\bibitem[{Esposito {et~al.}(2020)Esposito, Kalas, Fitzgerald, Millar-Blanchaer,
  Duchene, Patience, Hom, Perrin, Rosa, Chiang, Czekala, Macintosh, Graham,
  Ansdell, Arriaga, Bruzzone, Bulger, Chen, Cotten, Dong, Draper, Follette,
  Hung, Lopez, \& Matthews}]{Esposito:2020aa}
Esposito, T.~M., Kalas, P., Fitzgerald, M.~P., {et~al.} 2020, AJ, 60, 24,
  \dodoi{10.3847/1538-3881/ab9199}

\bibitem[{Foreman-Mackey(2016)}]{corner}
Foreman-Mackey, D. 2016, Journal of Open Source Software, 1, 24,
  \dodoi{10.21105/joss.00024}

\bibitem[{Foreman-Mackey {et~al.}(2013)Foreman-Mackey, Hogg, Lang, \&
  Goodman}]{Foreman-Mackey:2013aa}
Foreman-Mackey, D., Hogg, D.~W., Lang, D., \& Goodman, J. 2013, Publications of
  the ASP, 125, 306, \dodoi{10.1086/670067}

\bibitem[{{Gaia Collaboration} {et~al.}(2018){Gaia Collaboration}, Brown,
  Vallenari, Prusti, de~Bruijne, Babusiaux, \&
  Bailer-Jones}]{Collaboration:2018aa}
{Gaia Collaboration}, Brown, A. G.~A., Vallenari, A., {et~al.} 2018, A$\&$A,
  616, A1, \dodoi{10.1051/0004-6361/201833051}

\bibitem[{Heikamp \& Keller(2019)}]{Heikamp_19}
Heikamp, S., \& Keller, C.~U. 2019, A\&A, 627, A156,
  \dodoi{10.1051/0004-6361/201730557}

\bibitem[{Hughes {et~al.}(2018)Hughes, Duch\^{e}ne, \&
  Matthews}]{Hughes:2018aa}
Hughes, A.~M., Duch\^{e}ne, G., \& Matthews, B. 2018, ARA\&A, 56, 541-591,
  \dodoi{10.1146/annurev-astro-081817-052035}

\bibitem[{{Hung} {et~al.}(2016){Hung}, {Bruzzone}, {Millar-Blanchaer}, {Wang},
  {Arriaga}, {Metchev}, {Fitzgerald}, {Sivaramakrishnan}, \&
  {Perrin}}]{Hung:2016}
{Hung}, L.-W., {Bruzzone}, S., {Millar-Blanchaer}, M.~A., {et~al.} 2016, in
  \procspie, Vol. 9908, Society of Photo-Optical Instrumentation Engineers
  (SPIE) Conference Series, 99083A, \dodoi{10.1117/12.2233665}

\bibitem[{Hunter(2007)}]{Hunter07}
Hunter, J.~D. 2007, Computing in Science Engineering, 9, 90-95,
  \dodoi{10.1109/MCSE.2007.55}

\bibitem[{Kalas {et~al.}(2005)Kalas, Graham, \& Clampin}]{Kalas:2005aa}
Kalas, P., Graham, J.~R., \& Clampin, M. 2005, Nature, 435, 1067-1070,
  \dodoi{10.1038/nature03601}

\bibitem[{Kalas {et~al.}(2015)Kalas, Rajan, Wang, Millar-Blanchaer, Duchene,
  Chen, Fitzgerald, Dong, Graham, Patience, Macintosh, Murray-Clay, Matthews,
  Rameau, Marois, Chilcote, Rosa, Doyon, Draper, Lawler, Ammons, Arriaga,
  Bulger, Cotten, \& Follette}]{Kalas:2015aa}
Kalas, P.~G., Rajan, A., Wang, J.~J., {et~al.} 2015, ApJ, 814, 32,
  \dodoi{10.1088/0004-637X/814/1/32}

\bibitem[{Kennedy \& Wyatt(2010)}]{Kennedy:2010aa}
Kennedy, G.~M., \& Wyatt, M.~C. 2010, MNRAS, 405, 1253-1270,
  \dodoi{10.1111/j.1365-2966.2010.16528.x}

\bibitem[{Kennedy {et~al.}(2014)Kennedy, Wyatt, Kalas, Duch{\^e}ne, Sibthorpe,
  Lestrade, Matthews, \& Greaves}]{Kennedy:2014aa}
Kennedy, G.~M., Wyatt, M.~C., Kalas, P., {et~al.} 2014, MNRAS, 438, L96–L100,
  \dodoi{10.1093/mnrasl/slt168}

\bibitem[{Kenyon \& Bromley(2008)}]{Kenyon:2008aa}
Kenyon, S.~J., \& Bromley, B.~C. 2008, ApJ Letters, 179, 451,
  \dodoi{10.1088/0004-637X/690/2/L140}

\bibitem[{Kirchschlager \& Wolf(2013)}]{Kirchschlager:2013aa}
Kirchschlager, F., \& Wolf, S. 2013, A$\&$A, 552, A54,
  \dodoi{10.1051/0004-6361/201220486}

\bibitem[{Kral {et~al.}(2020)Kral, Matra, Kennedy, Marino, \&
  Wyatt}]{Kral:2020aa}
Kral, Q., Matra, L., Kennedy, G., Marino, S., \& Wyatt, M. 2020, MNRAS, 497,
  2811-2830, \dodoi{10.1093/mnras/staa2038}

\bibitem[{Krist {et~al.}(2005)Krist, Ardila, Golimowski, Clampin, Ford,
  Illingworth, Hartig, Bartko, Ben{\'{\i}}tez, Blakeslee, Bouwens, Bradley,
  Broadhurst, Brown, Burrows, Cheng, Cross, Demarco, Feldman, Franx, Goto,
  Gronwall, Holden, Homeier, Infante, Kimble, Lesser, Martel, Mei, Menanteau,
  Meurer, Miley, Motta, Postman, Rosati, Sirianni, Sparks, Tran, Tsvetanov,
  White, \& Zheng}]{Krist_2005}
Krist, J.~E., Ardila, D.~R., Golimowski, D.~A., {et~al.} 2005, ApJ, 129,
  1008-1017, \dodoi{10.1086/426755}

\bibitem[{Lagrange {et~al.}(2016)Lagrange, Langlois, Gratton, Maire, Milli,
  Olofsson, Vigan, Bailey, Mesa, Chauvin, Boccaletti, Galicher, Girard,
  Bonnefoy, Samland, Menard, Henning, Kenworthy, Thalmann, Beust, Beuzit,
  Brandner, Buenzli, Cheetham, \& Janson}]{Lagrange:2016aa}
Lagrange, A.~M., Langlois, M., Gratton, R., {et~al.} 2016, A$\&$A, 586, L8,
  \dodoi{10.1051/0004-6361/201527264}

\bibitem[{Lee \& Chiang(2016)}]{Lee:2016aa}
Lee, E.~J., \& Chiang, E. 2016, ApJ, 827, 125,
  \dodoi{10.3847/0004-637X/827/2/125}

\bibitem[{Li \& Greenberg(1998)}]{Li98}
Li, A., \& Greenberg, J.~M. 1998, A$\&$A, 338, 364

\bibitem[{Macintosh {et~al.}(2014)Macintosh, Graham, Ingraham, Konopacky,
  Marois, Perrin, Poyneer, Bauman, Barman, Burrows, Cardwell, Chilcote, Rosa,
  Dillon, Doyon, Dunn, Erikson, Fitzgerald, Gavel, Goodsell, Hartung, Hibon,
  Kalas, Larkin, \& Maire}]{Macintosh:2014aa}
Macintosh, B., Graham, J.~R., Ingraham, P., {et~al.} 2014, Proceedings of the
  National Academy of Science, 111, 12661, \dodoi{10.1073/pnas.1304215111}

\bibitem[{Macintosh {et~al.}(2018)Macintosh, Chilcote, Bailey, Rosa, Nielsen,
  Norton, Poyneer, Wang, Ruffio, Graham, Marois, Savransky, \&
  Veran}]{Macintosh:2018aa}
Macintosh, B., Chilcote, J.~K., Bailey, V.~P., {et~al.} 2018, SPIE, 10703,
  158-166, \dodoi{10.1117/12.2314253}

\bibitem[{Marshall {et~al.}(2017)Marshall, Maddison, Thilliez, Matthews,
  Wilner, Greaves, \& Holland}]{Marshall:2017aa}
Marshall, J.~P., Maddison, S.~T., Thilliez, E., {et~al.} 2017, MNRAS, 468,
  2719, \dodoi{10.1093/mnras/stx645}

\bibitem[{Matthews {et~al.}(2014)Matthews, Krivov, Wyatt, Bryden, \&
  Eiroa}]{Matthews:2014aa}
Matthews, B.~C., Krivov, A.~V., Wyatt, M.~C., Bryden, G., \& Eiroa, C. 2014,
  Protostars and Planets VI, 521,
  \dodoi{10.2458/azu_uapress_9780816531240-ch023}

\bibitem[{Millar-Blanchaer {et~al.}(2015)Millar-Blanchaer, Graham, Pueyo,
  Kalas, Dawson, Wang, Perrin, Moon, Macintosh, Ammons, Barman, Cardwell, Chen,
  Chiang, Chilcote, Cotten, Rosa, Draper, Dunn, Duch{\^e}ne, Esposito,
  Fitzgerald, Follette, Goodsell, \& Greenbaum}]{Millar-Blanchaer:2015aa}
Millar-Blanchaer, M.~A., Graham, J.~R., Pueyo, L., {et~al.} 2015, ApJ, 811, 18,
  \dodoi{10.1088/0004-637X/811/1/18}

\bibitem[{Milli {et~al.}(2017)Milli, Vigan, Mouillet, Lagrange, Augereau,
  Pinte, Mawet, Schmid, Boccaletti, Matrà, \& et~al.}]{Milli_2017}
Milli, J., Vigan, A., Mouillet, D., {et~al.} 2017, A\&A, 599, A108,
  \dodoi{10.1051/0004-6361/201527838}

\bibitem[{Min {et~al.}(2015)Min, Rab, Woitke, Dominik, \&
  M{\'e}nard}]{Min:2015aa}
Min, M., Rab, C., Woitke, P., Dominik, C., \& M{\'e}nard, F. 2015, A$\&$A, 585,
  A13, \dodoi{10.1051/0004-6361/201526048}

\bibitem[{Nesvold {et~al.}(2017)Nesvold, Naoz, \& Fitzgerald}]{Nesvold:2017aa}
Nesvold, E.~R., Naoz, S., \& Fitzgerald, M. 2017, ApJ, 837, L6,
  \dodoi{10.3847/2041-8213/aa61a7}

\bibitem[{Nguyen {et~al.}(2021)Nguyen, {De Rosa}, \& Kalas}]{Nguyen2020}
Nguyen, M.~M., {De Rosa}, R.~J., \& Kalas, P. 2021, AJ, 161, 22,
  \dodoi{10.3847/1538-3881/abc012}

\bibitem[{Oliphant(2006)}]{Oliphant_06}
Oliphant, T.~E. 2006, A Guide to NumPy, Vol. 1 (Spanish Fork, UT: Trelgol
  Publishing)

\bibitem[{Pawellek \& Krivov(2015)}]{Pawellek:2015aa}
Pawellek, N., \& Krivov, A.~V. 2015, MNRAS, 454, 3207–3221,
  \dodoi{10.1093/mnras/stv2142}

\bibitem[{Perez \& Granger(2007)}]{perez07}
Perez, F., \& Granger, B.~E. 2007, IPython: A System for Interactive Scientific
  Computing, 3, 21-29, 9, \dodoi{10.1109/MCSE.2007.53}

\bibitem[{Perrin {et~al.}(2014)Perrin, Maire, Ingraham, Savransky,
  Millar-Blanchaer, Wolff, Ruffio, Wang, Draper, Sadakuni, Marois, Rajan,
  Fitzgerald, Macintosh, Graham, Doyon, Larkin, Chilcote, Goodsell, Palmer,
  Labrie, Beaulieu, Rosa, Greenbaum, \& Hartung}]{Perrin:2014aa}
Perrin, M.~D., Maire, J., Ingraham, P., {et~al.} 2014, SPIE, 9147, 1168–1180,
  \dodoi{10.1117/12.2055246}

\bibitem[{Pinte {et~al.}(2006)Pinte, Menard, Duchene, \& Bastien}]{Pinte:aa}
Pinte, C., Menard, F., Duchene, G., \& Bastien, P. 2006, A$\&$A, 459, 797,
  \dodoi{10.1051/0004-6361:20053275}

\bibitem[{Rodet {et~al.}(2017)Rodet, Beust, Bonnefoy, Lagrange, Galli,
  Ducourant, \& Teixeira}]{Rodet:2017aa}
Rodet, L., Beust, H., Bonnefoy, M., {et~al.} 2017, A$\&$A, 602, A12,
  \dodoi{10.1051/0004-6361/201630269}

\bibitem[{Rodigas {et~al.}(2015)Rodigas, {Stark}, {Weinberger}, {Debes},
  {Hinz}, {Close}, \& {Chen}}]{Rodigas2015}
Rodigas, T.~J., {Stark}, C.~C., {Weinberger}, A., {et~al.} 2015, \apj, 798, 96,
  \dodoi{10.1088/0004-637X/798/2/96}

\bibitem[{Rodigas {et~al.}(2012)Rodigas, Hinz, Leisenring, Vaitheeswaran,
  Skemer, Skrutskie, Su, Bailey, Schneider, Close, Mannucci, Esposito,
  Arcidiacono, Pinna, Argomedo, Agapito, Apai, Bono, Boutsia, Briguglio, Brusa,
  Busoni, Cresci, Currie, \& Desidera}]{Rodigas:2012aa}
Rodigas, T.~J., Hinz, P.~M., Leisenring, J., {et~al.} 2012, ApJ, 752, 57,
  \dodoi{10.1088/0004-637X/752/1/57}

\bibitem[{Rouleau \& Martin(1991)}]{rouleau91}
Rouleau, F., \& Martin, P.~G. 1991, ApJ, 377, 526, \dodoi{10.1086/170382}

\bibitem[{Schneider {et~al.}(2014)Schneider, Grady, Hines, Stark, Debes,
  Carson, Kuchner, Perrin, Weinberger, Wisniewski, Silverstone, Jang-Condell,
  Henning, Woodgate, Serabyn, Moro-Martin, Tamura, Hinz, \&
  Rodigas}]{Schneider:2014aa}
Schneider, G., Grady, C.~A., Hines, D.~C., {et~al.} 2014, ApJ, 148, 59,
  \dodoi{10.1088/0004-6256/148/4/59}

\bibitem[{Strubbe \& Chiang(2006)}]{Strubbe:2006aa}
Strubbe, L.~E., \& Chiang, E.~I. 2006, ApJ, 648, 652-665,
  \dodoi{10.1086/505736}

\bibitem[{{The Astropy Collaboration} {et~al.}(2018){The Astropy
  Collaboration}, Price-Whelan, Sip{\H o}cz, G{\"u}nther, Lim, Crawford,
  Conseil, Shupe, Craig, Dencheva, Ginsburg, VanderPlas, Bradley,
  P{\'e}rez-Su{\'a}rez, de~Val-Borro, Aldcroft, Cruz, Robitaille, Tollerud,
  Ardelean, Babej, Bachetti, Bakanov, Bamford, \&
  Barentsen}]{Collaboration:2018ab}
{The Astropy Collaboration}, Price-Whelan, A.~M., Sip{\H o}cz, B.~M., {et~al.}
  2018, \dodoi{10.3847/1538-3881/aabc4f}

\bibitem[{Virtanen {et~al.}(2020)Virtanen, Gommers, Oliphant, Haberland, Reddy,
  Cournapeau, Burovski, \& Peterson}]{Virtanen_20}
Virtanen, P., Gommers, R., Oliphant, T.~E., {et~al.} 2020, Nature Methods, 17,
  261-272, \dodoi{10.1038/s41592-019-0686-2}

\bibitem[{Wang {et~al.}(2014)Wang, Rajan, Graham, Savransky, Ingraham,
  Ward-Duong, Patience, Rosa, Bulger, Sivaramakrishnan, Perrin, Thomas,
  Sadakuni, Greenbaum, Pueyo, Marois, Oppenheimer, Kalas, Cardwell, Goodsell,
  Hibon, Rantakyr{\"o}, \& with~the GPI~team}]{Wang:2014aa}
Wang, J.~J., Rajan, A., Graham, J.~R., {et~al.} 2014, SPIE, 152, 97,
  \dodoi{10.1117/12.2055753}

\bibitem[{Wyatt(2008)}]{Wyatt:2008aa}
Wyatt, M.~C. 2008, ARA\&A, 46, 339,
  \dodoi{10.1146/annurev.astro.45.051806.110525}

\end{thebibliography}
\bibliographystyle{aasjournal}



\end{document}